\documentclass[tightenlines,eqsecnum,floats,aps,amsmath,amssymb,nofootinbib,prd,showpacs]{revtex4-1}
\pdfoutput=1 
\usepackage{amsmath,amssymb,amsfonts}
\usepackage{graphicx}
\usepackage{enumerate} % advanced enumerate environment
\usepackage{colordvi} % for color text
\usepackage{bm}% bold math
\usepackage{epsfig}
\usepackage{float}
\usepackage{verbatim}
\usepackage{subfig}
\usepackage{natbib}
\newcommand{\bfv}{\mbox{\boldmath$v$}}
\newcommand{\bfx}{\mbox{\boldmath$x$}}
\newcommand{\bfk}{\mbox{\boldmath$k$}}
\newcommand{\bfp}{\mbox{\boldmath$p$}}
\newcommand{\bfq}{\mbox{\boldmath$q$}}
\newcommand{\bfr}{\mbox{\boldmath$r$}}

\begin{document}
\title{A Perturbative Approach to the Redshift Space Correlation Function: Beyond the Standard Model}

\vfill
\author{Benjamin Bose$^{1}$, Kazuya Koyama$^{1}$}
\bigskip
\affiliation{$^1$Institute of Cosmology \& Gravitation, University of Portsmouth,
Portsmouth, Hampshire, PO1 3FX, UK}
\bigskip
\vfill
\date{today}
\begin{abstract}   
We extend our previous redshift space power spectrum code to the redshift space correlation function. Here we focus on the Gaussian Streaming Model (GSM). Again, the code accommodates a wide range of modified gravity and dark energy models. For the non-linear real space correlation function used in the GSM we use the Fourier transform of the RegPT 1-loop matter power spectrum. We compare predictions of the GSM for a Vainshtein screened and Chameleon screened model as well as GR. These predictions are compared to the Fourier transform of the Taruya, Nishimichi and Saito (TNS) redshift space power spectrum model which is fit to N-body data. We find very good agreement between the Fourier transform of the TNS model and the GSM predictions, with $\leq 6\%$ deviations in the first two correlation function multipoles for all models for redshift space separations in $50$Mpc$/h \leq s \leq 180$Mpc/$h$. Excellent agreement is found in the differences between the modified gravity and GR multipole predictions for both approaches to the redshift space correlation function, highlighting their matched ability in picking up deviations from GR. We elucidate the timeliness of such non-standard templates at the dawn of stage-IV surveys and discuss necessary preparations and extensions needed for upcoming high quality data. 
\end{abstract}
\pacs{98.80.-k}
\maketitle
%%%%%%%%%%%%%%%%%%%%%%%%%%%%%%%%%%%%%%%%%%%%%%%%%%%%%%%
%%%%%%%%%%%%%%%%%%%%%%%%%%%%%%%%%%%%%%%%%%%%%%%%%%%%%%%
\section{Introduction}
%%%%%%%%%%%%%%%%%%%%%%%%%%%%%%%%%%%%%%%%%%%%%%%%%%%%%%%
%%%%%%%%%%%%%%%%%%%%%%%%%%%%%%%%%%%%%%%%%%%%%%%%%%%%%%%
Since its conception, Einstein's general relativity (GR) has enjoyed numerous successes, most recently the detection of long predicted gravitational waves \cite{Abbott:2016blz}. Despite this, with the discovery of cosmic acceleration \cite{Riess:1998cb,Perlmutter:1998np}, this model of gravity has needed to dramatically expand its `dark' sector to include dark energy in the form of a cosmological constant - $\Lambda$. This joins the long present dark matter (DM) component, putting the dark sector at around $95\%$ of the energy content of the universe \cite{Planck:2015xua}. Besides being non-interacting with light, making it very difficult to detect through conventional methods, the tiny value of the cosmological constant leads to the infamous fine tuning problem e.g. \cite{Weinberg:1988cp,Martin:2012bt}. By adopting GR, one is also lead to tensions between low redshift large scale structure data sets and the CMB \cite{Ade:2013lmv,Samushia:2012iq,Macaulay:2013swa,Heymans:2012gg,Riess:2011yx,Riess:2016jrr}. Motivated by these issues, alternatives to a pure $\Lambda$-CDM description of our universe have been investigated in great depth e.g.\cite{Hu:2007nk,Dvali:2000hr,Gomes:2013ema,Hassan:2011zd,deRham:2010ik,Maartens:2010ar,Pourtsidou:2016ico,Crisostomi:2016czh,Comelli:2011zm}. 
\newline
\newline
A basic perquisite of such modifications is to satisfy the many solar system tests that GR has passed over the years. But by modifying GR one generally introduces an additional degree of freedom usually cast in the form of a scalar field, which leads to an additional force on top of the GR-predicted one. This force is not observed in the local universe and so this naturally leads to the idea of screening, where the so called fifth force is shut off in areas of high density (See \cite{Koyama:2015vza,Sakstein:2015oqa,Clifton:2011jh} for reviews). Screening generally comes in two flavours, environmental screening where either the local density distribution effectively kills scalar field gradients sourcing the fifth force or it suppresses the coupling between the scalar and matter, and screening via new derivative interactions which modify the derivative structure of the Klein Gordon equation. The former can come in two forms, the Chameleon mechanism \cite{Hinterbichler:2010es} and the symmetron mechanism \cite{Khoury:2003aq,Khoury:2003rn,Hu:2007nk,Brax:2008hh}, while the latter is known as Vainshtein screening  \cite{Vainshtein:1972sx}. 
\newline
\newline
It is because of screening that the large scale structure (LSS) of the universe becomes such an attractive laboratory in which to test deviations from GR. At the relevant scales, the unscreened fifth force can work to produce a detectable signal of modified gravity (MG) \cite{Uzan:2000mz,Lue:2004rj,Ishak:2005zs,Knox:2005rg,Koyama:2006ef,Chiba:2007rb,Amendola:2007rr,Simpson:2012ra,Terukina:2012ji,Terukina:2013eqa,Yamamoto:2010ie,Jain:2007yk,Zhao:2008bn,Zhao:2009fn,Asaba:2013xql,Hellwing:2017pmj}. In particular the so called Redshift Space Distortions (RSD)  \cite{Kaiser:1987qv} in galaxy clustering promises to be a great candidate for testing MG. RSD is a matter clustering anisotropy that  comes from the non-linear mapping between real position and redshift space position. This mapping must account for the peculiar velocities of the clustering galaxies. In the presence of a fifth force, these peculiar velocities will be boosted giving us a signal in the observed anisotropy  \cite{Linder:2007nu,Guzzo:2008ac,Yamamoto:2008gr,Song:2008qt,Song:2010bk,Guzik:2009cm,Song:2010fg,Asaba:2013mxj,Hellwing:2014nma}.  
\newline
\newline
By comparing clustering statistics predicted by theory with the observed statistics coming from spectroscopic surveys such as the Baryon Oscillation Spectroscopic Survey within SDSS III (BOSS) \footnote{http://www.sdss3.org}, one can get a measurement of the growth of structure \cite{Samushia:2013yga,Gil-Marin:2015sqa,Sanchez:2016sas,Oka:2013cba,delaTorre:2013rpa}, typically coming in the form of the parameter combination $f\sigma_8$, where $f=d \ln(F_1)/ d\ln(a)$, with $F_1$ being the linear growth of structure and $a$ being the scale factor, while $\sigma_8$ normalises the linear power spectrum (see eg. \cite{Peebles1980}). Such a measurement can be used to constrain modified gravity parameters \cite{Dossett:2014oia,Johnson:2015aaa,Song:2015oza}, with the caveat that the limitations of the theoretical template applied to data are known. On this point, work has been done in determining the importance of using correctly modelled theoretical templates when comparing to data in order to obtain an unbiased measurement of growth and consequently MG parameters \cite{Taruya:2013my,Barreira:2016mg,Bose:2017myh}. A GR based modelling of RSD with additional degrees of freedom quantifying our ignorance (eg. tracer bias or velocity dispersions), has seemed to be a good approximation when placing general constraints on gravity. But as the precision of surveys increases, modelling gravity consistently becomes more and more important as any MG signal loses room within the error bars to hide. The same is also true as theory pushes its applicability to smaller scales, where the strongest statistical power comes from. 
\newline
\newline 
In \cite{Bose:2016qun} we presented a means for consistently constructing the redshift space power spectrum for an MG theory within the Horndeski class with a generalised potential assuming quasi static perturbations. The model for the redshift space power spectrum considered was the TNS model \cite{Taruya:2010mx}  built upon standard perturbation theory (SPT). This model has been shown to perform very well when compared to N-body simulations \cite{Nishimichi:2011jm,Taruya:2013my,Ishikawa:2013aea} and has consequently been employed in deriving constraints using survey data \cite{Beutler:2013yhm,Song:2015oza,Beutler:2016arn}. Part of its success at smaller scales is due to the introduction of a velocity dispersion parameter which is fit to data. The aim of \cite{Bose:2016qun} was to provide a means of applying a wide class of MG templates to data and in doing so bypass the question of model bias discussed in \cite{Taruya:2013my,Barreira:2016mg,Bose:2017myh}. This work moves in a similar vein. 
\newline
\newline
\newline
\newline
In this paper we move out of Fourier space and into configuration space by applying a similar approach as in \cite{Bose:2016qun} to the redshift space correlation function. On smaller scales, where the so called fingers-of-god effect and survey systematics become more dominant, a Fourier analysis of data becomes less ideal because $\bfk-$modes will not evolve independently. Large modes mixing with small ones affect our measurements/predictions at larger scales.
\newline
\newline
Traditionally the correlation function measured from surveys has been used for theory-data comparisons. Here we focus on the Gaussian streaming model (GSM) described in \cite{Reid:2011ar} which assumes a Gaussian form for the velocity probability distribution function. The GSM has been shown to fit N-body data to percent level on and below the Baryon Acoustic Oscillation (BAO) scale in the case of GR and has been widely applied to data \cite{Samushia:2012iq,Reid:2012sw,Wang:2013hwa,Satpathy:2016tct,Alam:2015qta}. Work has also been done in extending and modifying the model to achieve a better range of applicability \cite{Uhlemann:2015hqa,Bianchi:2016qen,Bianchi:2014kba}. When considering the first two multipole moments, the GSM can accurately probe down to scales $ s \geq 30 \mbox{Mpc}/h$ in the GR case when compared to N-body simulations for halos and matter and does very well in comparison to other redshift space correlation function models \cite{White:2014naa}. 
\newline
\newline
The TNS and GSM models are two of the most widely used theoretical templates applied to observational data. By having these two templates at one's disposal it is possible to make independent comparisons to data where observational systematics are different for each statistic. This will help beat down systematic uncertainty in growth measurements. Further, we compare the GSM predictions with the Fourier transform of the TNS power spectra. We expect the TNS transform to be the more accurate of the two because it makes use of the free parameter $\sigma_v$ to capture small scale dispersion effects whereas the pure GSM model contains no phenomenological ingredients. To merit the GSM approach, the TNS approach requires resummation techniques such as regularised perturbation theory (RegPT) \cite{Bernardeau:2008fa,Bernardeau:2011dp,Taruya:2012ut} to perform the Fourier transform. These techniques are currently under scrutiny in light of the effective field theory of large scale structure (EFToLSS)\cite{Baumann:2010tm,Carrasco:2012cv}. Suffice to say it is not well defined how one should treat small scale SPT divergences. These techniques are not needed in the GSM modelling. Of course one needs simulation measurements to see which approach does better, but their comparison has merit as a consistency check in their ability to model the observations as well as deviations from GR. This consistency is explored at and around the baryon acoustic oscillation (BAO) scale ($50\mbox{Mpc}/h \leq s \leq 180 \mbox{Mpc}/h$) in this paper. 
 \newline
 \newline
%EDIT: included above paragraph
% comment on its importance for MG detection/constraints
%comment in the conclusion on moments of velocity PDF on small scales - Hellwing et al 2014 and Fontanot et al 2013
This paper is organised as follows: In Sec.II we review the SPT building blocks for the construction of the GSM for generalised gravitational theory templates.  This is followed by a review of the TNS power spectrum which we will use to compare with our GSM predictions. We conclude this section by going over the GSM and its ingredients in terms of the generalised perturbation kernels.  In Sec.III we present the GSM predictions. We validate the predictions  by comparing them with the Fourier transform of TNS predictions fitted to three sets of N-body data, each run under a different model of gravity. Finally, we summarise our results, review our pipeline, its successes and shortcomings, and highlight future work in Sec.IV.  
%%%%%%%%%%%%%%%%%%%%%%%%%%%%%%%%%%%%%%%%%%%%%%%%%%%%%%%
%%%%%%%%%%%%%%%%%%%%%%%%%%%%%%%%%%%%%%%%%%%%%%%%%%%%%%%
\section{Theory}
\subsection{Evolution equations for the perturbations}
%%%%%%%%%%%%%%%%%%%%%%%%%%%%%%%%%%%%%%%%%%%%%%%%%%%%%%%
%%%%%%%%%%%%%%%%%%%%%%%%%%%%%%%%%%%%%%%%%%%%%%%%%%%%%%%
We begin by considering perturbations in homogeneous and isotropic expanding background. This is strongly supported by various experiments, most notably the Planck mission \footnote{http://sci.esa.int/planck/}. These can be described by the perturbed Friedman-Robertson-Walker (FRW) metric. In the Newtonian gauge this is given by 
%%%%%%%%%%%%%%%%%%%%%%%%%%%%%%%%%%%%%%%%%%%%%%%%%%%%%%%%%%%%%%%%%%%%%%
\begin{equation}
ds^2=-(1+2\Phi)dt^2+a^2(1-2\Psi)\delta_{ij}dx^idx^j.
\end{equation}
%%%%%%%%%%%%%%%%%%%%%%%%%%%%%%%%%%%%%%%%%%%%%%%%%%%%%%%%%%%%%%%%%%%%%%
We will assume the matter content of the universe is well described by a perfect fluid and will work on the evolution of density fluctuations inside the Hubble horizon. We work on sub-horizon scales where we can use the quasi-static approximation and neglect time derivatives of the perturbed quantities compared with the spatial derivatives. It is worth pointing out that the validity of this approximation weakens in certain cases.  The large distance modification of gravity which is invoked to explain cosmic acceleration introduces a new scalar degree of freedom. The quasi static approximation generally holds as long as the sound speed of this degree of freedom is marginally larger than the speed of light \cite{Sawicki:2015zya} within the scales considered. This is model and scale dependant and one should be aware of its validity when applying the treatment described here. 
\newline
\newline
In this paper we work in the Jordan frame. The evolution equations for matter perturbations are obtained from the conservation of the energy momentum tensor. Before shell crossing and assuming no vorticity in the velocity field, a safe assumption at large scales and late times, the evolution equations can be expressed in Fourier space as 
%%%%%%%%%%%%%%%%%%%%%%%%%%%%%%%%%%%%%%%%%%%%%%%%%%%%%%%%%%%%
\begin{eqnarray}
&&a \frac{\partial \delta(\bfk)}{\partial a}+\theta(\bfk) =-
\int\frac{d^3\bfk_1d^3\bfk_2}{(2\pi)^3}\delta_{\rm D}(\bfk-\bfk_1-\bfk_2)
\alpha(\bfk_1,\bfk_2)\,\theta(\bfk_1)\delta(\bfk_2),
\label{eq:Perturb1}\\
&& a \frac{\partial \theta(\bfk)}{\partial a}+
\left(2+\frac{a H'}{H}\right)\theta(\bfk)
-\left(\frac{k}{a\,H}\right)^2\,\Phi(\bfk)=
-\frac{1}{2}\int\frac{d^3\bfk_1d^3\bfk_2}{(2\pi)^3}
\delta_{\rm D}(\bfk-\bfk_1-\bfk_2)
\beta(\bfk_1,\bfk_2)\,\theta(\bfk_1)\theta(\bfk_2),
\label{eq:Perturb2}
\end{eqnarray}
%%%%%%%%%%%%%%%%%%%%%%%%%%%%%%%%%%%%%%%%%%%%%%%%%%%%%%%%%%%%%%%%%%%%%%
where the prime denotes a scale factor derivative, $\delta$ is the density contrast representing an over or under density in the background density field and $\theta$ is the velocity divergence expressed in terms of the peculiar velocity field $\bfv_p({\bfx})$ as $\theta({\bf x})= \frac{\nabla\cdot \bfv_p({\bfx})}{aH(a) f}$. $H(a)$ is the Hubble factor and $f=d \ln(F_1)/ d\ln(a)$, with $F_1$ being the linear growth of structure defined below. The kernels in the Fourier integrals, $\alpha$  and $\beta$, are given by
%%%%%%%%%%%%%%%%%%%%%%%%%%%%%%%%%%%%%%%%%%%%%%%%%%%%%%%%%%%%%%%%%%%%%%
\begin{eqnarray}
\alpha(\bfk_1,\bfk_2)=1+\frac{\bfk_1\cdot\bfk_2}{|\bfk_1|^2},
\quad\quad
\beta(\bfk_1,\bfk_2)=
\frac{(\bfk_1\cdot\bfk_2)\left|\bfk_1+\bfk_2\right|^2}{|\bfk_1|^2|\bfk_2|^2}.
\label{alphabeta}
\end{eqnarray}
%%%%%%%%%%%%%%%%%%%%%%%%%%%%%%%%%%%%%%%%%%%%%%%%%%%%%%%%%%%%%%%%%%%%%%
The assumption of perturbation theory is that the non-linear density and velocity perturbations can be written out as a perturbative expansion of increasing order in the linear perturbations (See Ref.\cite{Bernardeau:2001qr} for a review).  Once we assume this, we can solve eq.(\ref{eq:Perturb1}) and eq.(\ref{eq:Perturb2}) order by order. For our needs we will expand $\delta$ and $\theta$ up to the third order. 
\newline
\newline
Gravity enters the evolution via the Newtonian potential $\Phi$ in the Euler equation (eq.(\ref{eq:Perturb2})) and its non-linear relation to the matter perturbations  is governed by the Poisson equation \cite{Koyama:2009me}.
%%%%%%%%%%%%%%%%%%%%%%%%%%%%%%%%%%%%%%%%%%%%%%%%%%%%%%%%%%%%%%%%%%%%%%
\begin{equation}
-\left(\frac{k}{a H}\right)^2\Phi=
\frac{3 \Omega_m(a)}{2} \mu(k,a)\,\delta(\bfk) + S(\bfk),
\label{eq:poisson1}
\end{equation}
%%%%%%%%%%%%%%%%%%%%%%%%%%%%%%%%%%%%%%%%%%%%%%%%%%%%%%%%%%%%%%%%%%%%%%
where $\Omega_m(a) = 8 \pi G \rho_m/3 H^2$ and the function $S(\bfk)$ is the non-linear source, term up to the third order given by
%%%%%%%%%%%%%%%%%%%%%%%%%%%%%%%%%%%%%%%%%%%%%%%%%%%%%%%%%%%%%%%%%%%%%%
\begin{eqnarray}
S(\bfk)&=&
\int\frac{d^3\bfk_1d^3\bfk_2}{(2\pi)^3}\,
\delta_{\rm D}(\bfk-\bfk_{12}) \gamma_2(\bfk, \bfk_1, \bfk_2;a)
\delta(\bfk_1)\,\delta(\bfk_2),
\nonumber\\
&& + 
\int\frac{d^3\bfk_1d^3\bfk_2d^3\bfk_3}{(2\pi)^6}
\delta_{\rm D}(\bfk-\bfk_{123})
\gamma_3( \bfk, \bfk_1, \bfk_2, \bfk_3;a)
\delta(\bfk_1)\,\delta(\bfk_2)\,\delta(\bfk_3),
\label{eq:Perturb3}
\end{eqnarray}
%%%%%%%%%%%%%%%%%%%%%%%%%%%%%%%%%%%%%%%%%%%%%%%%%%%%%%%%%%%%%%%%%%%%%%
where $\gamma_2( \bfk, \bfk_1, \bfk_2; a)$  and $\gamma_3(\bfk, \bfk_1, \bfk_2, \bfk_3;a)$ are symmetric under the exchange of $\bfk_i$. Expressions for these functions are derived under the quasi static Klein Gordon equation for the additional scalar degree of freedom typical in MG theories. In the case of GR $\gamma_2( \bfk, \bfk_1, \bfk_2; a) = \gamma_3(\bfk, \bfk_1, \bfk_2, \bfk_3;a) = 0$ and  $\mu(k,a)=1$.  In \cite{Bose:2016qun} we specify the form of these functions for the Vainshtein screened DGP model of gravity \cite{Dvali:2000hr} and for the chameleon screened Hu-Sawicki form of $f(R)$ gravity \cite{Hu:2007nk}. 
\newline
\newline
As said, we solve eq.(\ref{eq:Perturb1}) and eq.(\ref{eq:Perturb2}) perturbatively for \emph{n}-th order kernels $F_n$ and $G_n$ which give the \emph{n}-th order solutions 
\begin{align} 
\delta_n(\boldsymbol{k} ; a) &= \int d^3\boldsymbol{k}_1...d^3 \boldsymbol{k}_n \delta_D(\boldsymbol{k}-\boldsymbol{k}_{1...n}) F_n(\boldsymbol{k}_1,...,\boldsymbol{k}_n ; a) \delta_0(\boldsymbol{k}_1)...\delta_0(\boldsymbol{k}_n), \label{nth1} \\ 
\theta_n(\boldsymbol{k}; a) &= \int d^3\boldsymbol{k}_1...d^3 \boldsymbol{k}_n \delta_D(\boldsymbol{k}-\boldsymbol{k}_{1...n}) G_n(\boldsymbol{k}_1,...,\boldsymbol{k}_n; a) \delta_0(\boldsymbol{k}_1)...\delta_0(\boldsymbol{k}_2), \label{nth2}
\end{align}
where $\boldsymbol{k}_{1...n} = \boldsymbol{k}_1 + ...+ \boldsymbol{k}_n$. In the following we will omit the scale factor dependence of the kernels to simplify the expressions. Using the numerical algorithm described in \cite{Taruya:2016jdt} we can do this for all orders in a general way. Using the perturbations up to third order we can construct the so called 1-loop power spectrum
\begin{equation}
P^{1-{\rm loop}}_{ij}(k) = P_{0}(k) + P^{22}_{ij}(k) + P^{13}_{ij}(k),
\label{loopps}
\end{equation}
where $P_{0}(k)$ is the linear power spectrum defined as 
\begin{equation}
\langle \delta_0(\bfk) \delta_0(\bfk')\rangle =
(2\pi)^3\delta_{\rm D}(\bfk+\bfk')\,P_{0}(k),
\end{equation} 
$\langle ... \rangle$ denoting an ensemble average. The higher order terms are given by
\begin{align}
\langle g_i^{2}(\bfk) g_{j}^2(\bfk')\rangle &=
(2\pi)^3\delta_{\rm D}(\bfk+\bfk')\,P_{ij}^{22}(k), \label{eq:psconstraint0} \\
\langle g_i^{1}(\bfk) g_{j}^3(\bfk')
+g_i^{3}(\bfk) g_{j}^1(\bfk') \rangle &=
(2\pi)^3\delta_{\rm D}(\bfk+\bfk')\,P_{ij}^{13}(k),
\label{eq:psconstraint1}
\end{align}
where $g^i_1 = \delta_i$ and $g^i_2= \theta_i$.  The power spectrum is a Fourier space measure of the correlation between the fields and the higher order or 1-loop terms give the first order in non-linearity on top of the linear power spectrum. The inclusion of loop terms has been shown to improve the prediction of theory \cite{Jeong:2006xd}, an improvement more pronounced at higher redshift \cite{Carlson:2009it}. Despite this, the loop expansion of the power spectrum is known to have divergent behaviour at small scales making the benefits of the 1-loop terms only enjoyable within a restricted range of scales. Further this bad behaviour makes it difficult to move out of Fourier space which involves an integral over UV scales. In the next section we describe the RegPT treatment which allows us to do the Fourier transform safely. This treatment will be employed to Fourier transform the best fit TNS power spectrum in Sec.III which we compare to the GSM. We proceed by reviewing the TNS model of RSD. 
%%%%%%%%%%%%%%%%%%%%%%%%%%%%%%%%%%%%%%%%

\subsection{Modelling RSD in Fourier Space : TNS}
As we mentioned in the introduction, the anisotropy of galaxy clustering provides a very promising means of testing gravity.  The anisotropy arises from the non-linear mapping between real and redshift space and it is because of the mapping's non-linear nature that makes modelling RSD complex. A model of the effect was first given by Kaiser \cite{Kaiser:1987qv} which captures the coherent, linear peculiar infalling motion of galaxies in a cluster
\begin{equation}
P_{K}^S(k,\mu;a) = (1+f\mu^2)^2P_{\delta \delta}(k;a),
\label{linkais}
\end{equation}
where $\mu$ \footnote{The use of $\mu$ here should not be confused with the function $\mu(k;a)$ which will always include its arguments.} is the cosine of the angle between the line of sight and $\bfk$, $f = d \ln{F_1}/d\ln{a}$ is the logarithmic growth rate and $P_{\delta \delta}(k)$ is the linear matter power spectrum. The model does not account for the small scale damping effect of incoherent virialised motion - the fingers-of-god effect. Many authors accounted for this effect via a phenomenological pre factor term, usually taking the form of an exponential or Gaussian \cite{Scoccimarro:2004tg,Percival:2008sh,Cole:1994wf,Peacock:1993xg,Park:1994fa,Ballinger:1996cd,Magira:1999bn}. 
\newline
\newline
For robust tests of gravity a move beyond linear models is needed. As mentioned in the introduction, the non-linear TNS model of RSD has proved its merit against N-body and survey data. It is derived partly perturbatively with the small scale fingers-of-god effect being treated phenomenologically through a damping factor. The derivation can be followed in \cite{Taruya:2010mx}, but we simply present it here
 \begin{equation}
 P^S(k,\mu) = \mbox{D}_{\mbox{FoG}} (k\mu \sigma_v) \{ P^{1-{\rm loop}}_{\delta \delta} (k) - 2  \mu^2 P^{1-{\rm loop}}_{\delta \theta}(k) + \mu^4 P^{1-{\rm loop}}_{\theta \theta} (k) + A(k,\mu) + B(k,\mu) \}. 
 \label{redshiftps}
 \end{equation}
\noindent The A and B terms account for higher-order interactions between the density and velocity fields and are given by 
 \begin{equation}
 A(k,\mu)=  -(k \mu) \int d^3 \boldsymbol{k'} \left[  \frac{k_z '}{k'^2} B_\sigma(\boldsymbol{k'},\boldsymbol{k}-\boldsymbol{k'},-\boldsymbol{k}) +  \frac{k\mu-k_z'}{|\bfk-\bfk'^2|} B_\sigma(\boldsymbol{k}-\boldsymbol{k'}, \boldsymbol{k'},-\boldsymbol{k}) \right],
 \label{Aterm2}
 \end{equation}
 \begin{equation}
 B(k,\mu)= (k \mu)^2 \int d^3\boldsymbol{k'} F(\boldsymbol{k'}) F(\boldsymbol{k}-\boldsymbol{k'}),
 \label{Bterm}
 \end{equation}
 where
 \begin{equation}
 F(\boldsymbol{k}) = \frac{k_z}{k^2}\left[P_{\delta \theta} (k) - \frac{k_z^2}{k^2}P_{\theta \theta} (k) \right],
 \end{equation}
\noindent where the power spectra here are calculated at linear order. The cross bispectrum $B_\sigma$ is given by
 \begin{equation}
 \delta_D(\boldsymbol{k}_1+ \boldsymbol{k}_2+ \boldsymbol{k}_3)B_\sigma( \boldsymbol{k}_1,\boldsymbol{k}_2,\boldsymbol{k}_3) = \langle \theta(\boldsymbol{k}_1)\Big\{ \delta(\boldsymbol{k}_2) - \frac{k_{2z}^2}{k_2^2} \theta(\boldsymbol{k}_2)\Big\}\Big\{ \delta(\boldsymbol{k}_3) - \frac{k_{3z}^2}{k_3^2} \theta(\boldsymbol{k}_3)\Big\}\rangle,
 \label{lcdmbi}
 \end{equation}
\newline 
We choose an exponential form for  the fingers-of-god damping factor $D_{FoG}(k\mu\sigma_v)= \exp{(-k^2 \mu^2 \sigma_v^2)}$, where $\sigma_v$ is treated as a free parameter quantifying the dispersion in velocities (expressed in units Mpc/$h$) \cite{Peacock1992}. 
\newline
\newline
eq.(\ref{redshiftps}) gives very good predictions in the mildly non-linear regime but due to convergence problems of SPT a Fourier transform of the expression is difficult. Since we are concerned with configuration space statistics in this paper we will adopt the RegPT treatment for the power spectra components as well as for the A and B correction terms. This treatment essentially damps the divergences of the loop terms allowing an easy transform to configuration space.
\subsubsection{RegPT}
We go over the terms used to construct eq.(\ref{redshiftps}) using the RegPT treatment described in e.g \cite{Taruya:2014faa}. This treatment is based on expanding the multipoint propagators which contain the entire non-perturbative nature of the field's evolution \cite{Bernardeau:2008fa}. These propagators can be analytically described in terms of the perturbative kernels and allow for the construction of a quasi non-linear power spectrum and correlation function which show excellent agreement with N-body data in real and redshift space as well as for models other than GR \cite{Taruya:2012ut,Taruya:2013my,Taruya:2014faa}. 
\newline
\newline
In terms of the multipoint propagators $\Gamma_a^{(n)}$ the RegPT 1-loop power spectrum is given by 
\begin{align}
P_{bc}(k;a) =& \Gamma^{(1)}_b(k;a)\Gamma_c^{(1)}(k;a)P_0(k) \nonumber \\
 & + 2 \int \frac{d^3 \bfq}{(2\pi)^3} \Gamma_b^{(2)}(\bfq,\bfk-\bfq;a)\Gamma_c^{(2)}(\bfq,\bfk-\bfq;a)P_0(q)P_0(|\bfk-\bfq|),
 \label{regptloop}
\end{align}
where $P_0$ is the initial linear matter power spectrum and $b,c \in \{\delta ,\theta\}$.The propagators are given in terms of the perturbative kernels by  

\begin{align}
\Gamma^{(1)}_b(k;a) = & \left[ J_b^{(1)}(k;a)\Big\{ 1+ \frac{k^2\sigma_d^2}{2}\Big\}  \right. \nonumber \\
& \left. + 3\int \frac{ d^3\bfq}{(2\pi)^3} J_b^{(3)}(\bfk,\bfq,-\bfq;a)P_0(q)\right] e^{-k^2\sigma_d^2/2}, \\ 
\Gamma^{(2)}_b(\bfq,\bfk-\bfq;a)  = & J_b^{(2)}(\bfq,\bfk-\bfq;a)e^{-k^2\sigma_d^2/2}, 
\end{align}
where $J_b^{(n)} = (F_n,G_n)$. $\sigma_d^2$ is the dispersion of the linear displacement field given by 
\begin{equation}
\sigma_d^2(k) = \int^{k/2}_0 \frac{dq}{6\pi^2} F_1(q;a)^2P_0(q).
\end{equation}
This accounts for the loop terms in eq.(\ref{redshiftps}). The $A$ and $B$ correction terms are evaluated at tree-level since they are treated as next to leading order and are given in terms of the propagators as 
\begin{align}
P_{bc,tree}(k;a) &= \Gamma_b^{(1)}(k;a)\Gamma_c^{(1)}(k;a)P_0(k), \\
B_{bcd,tree}(\bfk_1,\bfk_2,\bfk_3) &= 2 \Gamma_b^{(2)}(\bfk_2,\bfk_3;a)\Gamma^{(1)}_c(k_2;a),\Gamma_d^{(1)}(k_3;a) P_0(k_2)P_0(k_3) + ({\rm cyc. perm})
\end{align}
where now the propagators are evaluated at tree level too 
\begin{equation}
\Gamma_{b,tree}^{(n)}(\bfk_1, ... \bfk_n;a) = J_a^{(n)}(\bfk_1 ... \bfk_n;a)e^{-k_{1...n}^2\sigma_d^2/2},
\end{equation}
where $k_{1...n} = |\bfk_1 + ... + \bfk_n|$. We end by noting that the generality of these expressions applies to all MG models within the framework described in \cite{Bose:2016qun} although corrections may be incurred in some models as discussed in \cite{Taruya:2014faa}. For the models treated in this paper the corrections are negligible at the scales of interest. Finally, we will conclude this section by reviewing the GSM model for the redshift space correlation function. 
\newpage
%%%%%%%%%%%%%%%%%%%%%%%%%%%%%%%%

\subsection{Modelling RSD in Configuration Space: GSM} 
In \cite{Fisher:1994ks} the first study of the relation between eq.(\ref{linkais}) and the redshift space correlation function was made. Working in the linear regime, the authors developed the so called linear streaming model (LSM) which uses the mean infall velocity between pairs ($v_{12}$)  and the velocity dispersion along the line of sight (LOS) ($\sigma_{12}^2 $) as ingredients connecting theory to the RSD phenomenon in the linear regime. Specifically, it is the scale dependence of $v_{12}$ and $\sigma_{12}^2 $ which drives the distribution of galaxies away from isotropy. By assuming a Gaussian form for the joint density-velocity distribution, the LSM for biased tracers is given by 

\begin{equation}
1+\xi^s_{\rm LSM} (r_\sigma, r_\pi) = \int G(r,y) e^{-[r_\pi - y]^2/2\sigma_{\rm 12,lin}^2(r,\mu)} \frac{dy}{\sqrt{2\pi\sigma_{\rm 12,lin}^2(r,\mu)}}, 
\end{equation}
where 
\begin{equation}
G(r,y) = \left[1 + \xi^r_L(r) + \frac{y}{r} \frac{(r_\pi -y)v_{\rm 12,lin}(r)}{\sigma_{\rm 12,lin}^2(y)} - \frac{1}{4}\frac{y^2}{r^2}\frac{v_{\rm 12,lin}^2(r)}{\sigma_{\rm 12,lin}^2(y)}\left(1-\frac{(r_\pi-y)^2}{\sigma_{\rm 12,lin}^2(y)}\right) \right],
\end{equation} 
The variables are as follows: $r_\pi$ and $y$ are the separations parallel to the LOS of matter particles in redshift and real space respectively, $r_\sigma$  is the separation perpendicular to the LOS. $\xi^r_L$ is the linear real space galaxy correlation function determined by Fourier transforming the linear power spectrum, $v_{\rm 12,lin} = \langle \delta(\bfx) \bfv(\bfx')\rangle $  is the linear mean infall velocity of a particle pair with real space separation $r = \sqrt{y^2 + r_\sigma^2}$ and $\sigma_{\rm 12,lin}^2(r) = \langle (\bfv_{\rm LOS}(\bfx) -\bfv_{\rm LOS}(\bfx') )^2 \rangle$ is the linear velocity dispersion. The linear predictions for these are given below in terms of the generalised 1st order perturbative kernels $(F_1,G_1)$
\begin{equation}
\bfv_{\rm 12,lin}(r) = v_{\rm 12,lin}\hat{\bfr} = \frac{b\hat{\bfr}}{\pi^2} \int dk k  j_1(kr) G_1(k;a) F_1(k;a) P_0(k),
\label{linv12}
\end{equation}
 where $j_1(k)$ is the 1st order spherical Bessel function and $b$ is the linear bias factor.
 \begin{equation} 
 \sigma_{\rm 12,lin}^2(r, \mu^2) = 2\left[\sigma_1^2 - \frac{1}{2\pi^2}\int dk G_1(k;a) \mathcal{J}(kr,\mu^2) P_0(k)  \right],
 \label{lins12}
 \end{equation}
 where $\sigma_1^2 =\frac{1}{3}\langle \bfv(\bfx) \cdot \bfv(\bfx) \rangle$ is the 1-dimensional velocity dispersion and 
  \begin{equation} 
 \mathcal{J}(kr,\mu^2) = \mu^2\left(j_0(kr)-\frac{2j_1(kr)}{kr}\right) + (1-\mu^2)\frac{j_1(kr)}{kr}.
 \label{mybessel}
 \end{equation}
\newline
\newline
Moving away from the linear regime, we consider the non-linear redshift space correlation function developed in \cite{Reid:2011ar}, known as the Gaussian streaming model from its core assumption that the matter's pairwise velocity probability distribution is of a Gaussian form. This is given by
\begin{equation}
1+\xi^s_{\rm GSM}(r_\sigma, r_\pi) = \int [1+\xi^r(r)] e^{-[r_\pi - y - \mu v_{12}(r)]^2/2\sigma_{12}^2(r,\mu)} \frac{dy}{\sqrt{2\pi\sigma_{12}^2(r,\mu)}}.
\label{redshiftcor}
\end{equation} 
$\xi^r$ is the non-linear real space correlation function, $v_{12}(r)$ is the non-linear mean infall velocity of a particle pair and $\sigma_{12}^2(r,\mu)$ is the non-linear, non-isotropic velocity dispersion. Note for biased tracers we must include tracer bias in $\xi^r$. At linear bias level, this is given as $b^2\xi^r$, where $b$ is the linear bias factor.
\newline
\newline
In \cite{Reid:2011ar} the authors use the Lagrangian perturbation theory (LPT) model of \cite{Matsubara:2008wx} for the real space correlation function. Here we use a RegPT prescription where our real space correlation function is produced by Fourier transforming the RegPT 1-loop matter power spectrum
\begin{equation}
\xi^r(r) = \int \frac{d^3k}{(2\pi)^3} e^{i \footnotesize{\bfk\cdot \bfx}} P^{\rm 1-loop,RegPT}_{\delta \delta} (k). 
\label{pregab}
\end{equation}
\newpage
\noindent Because of RegPT's damping of the power spectrum at small scales, we can do the above integral without having to worry about SPT divergences.  As discussed in a previous section, $P^{{\rm1-loop}}_{ab}(k)$ can be readily constructed for general models of gravity. 
\newline
\newline
Finally, the mean infall velocity and velocity dispersion are given by correlations between the density field and the velocity field. Using a perturbative treatment of the fields we can derive expressions for these ingredients for general models of gravity in the linear and quasi non-linear regime. Here we give expressions for $v_{12}$ and $\sigma_{12}^2$ appearing in eq.(\ref{redshiftcor}) in terms of the generalised kernels $(F_n,G_n)$ (see eq.(\ref{nth1}) and eq.(\ref{nth2})). In the case of GR, using the Einstein de Sitter approximation for the kernels, one can follow Appendices A1 and A2 of \cite{Reid:2011ar}.  

\subsubsection{Mean Infall Velocity $v_{12}(r)$}
The mean infall velocity arrises from correlating the density field with the velocity. In terms of these correlations we can write (eq.(27) of \cite{Reid:2011ar}) 
\begin{equation}
[1+b^2\xi^r(r)] v_{12}(r)\hat{r} = 2b\langle \delta_1(\bfx)\bfv_1(\bfx+\bfr)\rangle + 2b\sum_{i>0} \langle \delta_i(\bfx)\bfv_{4-i}(\bfx+\bfr)\rangle +2b^2\sum_{i,j>0} \langle \delta_i(\bfx) \delta_j(\bfx+\bfr)\bfv_{4-i-j}(\bfx+\bfr) \rangle,
\end{equation}
where we have included $b$, the linear bias factor. $\xi^r$ is the matter correlation function and $\bfv$ is the velocity field perturbation. The correlations in the above expression up to 2nd order in the linear power spectrum are given below. 
\begin{equation}
 2b \left( \langle \delta_1(\bfx)\bfv_1(\bfx+\bfr)\rangle + \sum_{i>0} \langle \delta_i(\bfx)\bfv_{4-i}(\bfx+\bfr)\rangle \right) = \hat{\bfr} \frac{b}{\pi^2} \int dk k P_{\delta \theta}^{{\rm 1- loop}}(k; a) j_1(kr),
 \end{equation}
where $P_{\delta \theta}^{{\rm 1- loop}}$ is given by eq.(\ref{regptloop}). The last term has three contributions at 1-loop order, $i,j=(1,1),(1,2),(2,1)$.  The contribution from the first two of these (A5 of \cite{Reid:2011ar}) is given as
\begin{align} 
2b^2 \left( \langle \delta_1(\bfx) \delta_1(\bfx+\bfr)\bfv_{2}(\bfx+\bfr) \rangle + \langle \delta_1(\bfx) \delta_2( \right. &\left.\bfx+\bfr)\bfv_{1}(\bfx+\bfr) \rangle \right) = \nonumber \\
   \frac{b^2\hat{\bfr}}{2\pi^4} \int_0^\infty dk dy \int_{-1}^1 dx k^4 y j_1(kr) P_0(k)P_0(ky) & \left(  \frac{F_1(k;a)F_1(ky;a) G_2(ky,k,-x;a) y(1-yx)}{1+y^2-2yx} \right. \nonumber \\ &  + G_1(ky;a)F_1(k;a)F_2(ky,k,-x;a)x \bigg), 
\end{align}
and the (2,1) contribution (A6 of \cite{Reid:2011ar}) is given as 
\begin{align}
 2b^2  \langle \delta_2(\bfx) \delta_1(\bfx+\bfr)\bfv_{1}(\bfx+\bfr) \rangle = & \frac{b^2\hat{\bfr}}{2\pi^4} \int_0^\infty dk dy  \int_{-1}^1 dx k^4 y x j_1(kr) P_0(ky)P_0(k\sqrt{1+y^2-2yx}) \nonumber \\ & \times  F_1(k\sqrt{1+y^2-2yx};a)G_1(ky;a)F_2(ky,k\sqrt{1+y^2-2yx}, u ;a), 
\end{align}
where we have written the kernels in terms of the integrated vector's magnitudes and angle between them:  $|\bfk| = k$, $|\bfp| =  ky$ and $|\bfk - \bfp| = k\sqrt{1+y^2-2yx}$ with $x = \hat{\bfk}\cdot\hat{\bfp}$ and $u = \hat{(\bfk-\bfp)}\cdot \hat{\bfp}$. This notation is used for the velocity dispersion expressions below.

\subsubsection{Velocity Dispersion $\sigma_{12}^2(r,\mu^2)$}
The velocity dispersion depends on both the separation of the pair $r$ and the angle the separation vector $\bfr$ makes with the LOS, $\phi_{lr}$, expressed through the argument $\mu^2 = {\rm cos}^2 (\phi_{lr})$. One can combine the perpendicular and parallel components of $\sigma_{12}^2$ to get the expression  (eq.(29) to eq.(32) of \cite{Reid:2011ar}) 
\begin{align}
[1+b^2\xi^r(r)] \sigma_{12}^2(r,\mu^2) &= 2 \left( \langle (v^{\ell}(\bfx))^2\rangle - \langle v^{\ell}(\bfx)  v^{\ell}(\bfx+\bfr) \rangle \right)  + 2b \langle \delta(\bfx)( v^{\ell}(\bfx))^2\rangle  \nonumber \\
& + 2b \left[\langle \delta(\bfx) (v^{\ell}(\bfx+\bfr))^2 - 2 \langle \delta(\bfx)  v^{\ell}(\bfx) v^{\ell}(\bfx+\bfr) \rangle \right] \nonumber \\
& + 2b^2 \left[ \langle \delta(\bfx) \delta(\bfx + \bfr)  (v^{\ell}(\bfx))^2 \rangle - \langle \delta(\bfx) \delta(\bfx + \bfr)  v^{\ell}(\bfx)  v^{\ell}(\bfx + \bfr) \rangle \right], 
\label{sig12m}
\end{align} 
where $\ell$ denotes the component of $\bfv$ along the LOS. We give these component by component below. 
\begin{align} 
 2 \langle (v^{\ell}(\bfx))^2\rangle = 6 \sigma_1^2 = &  \frac{1}{3\pi^2} \int dk \frac{j_1(kr)}{kr} G_1(k;a)^2 P_0(k), \\ 
 - 2 \langle v^{\ell}(\bfx)  v^{\ell}(\bfx+\bfr) \rangle = & -\frac{1}{\pi^2} \int dk P_{\theta \theta}^{\rm 1-loop} (k;a) \mathcal{J}(kr,\mu^2),  
 \end{align}
 where again $P_{\theta \theta}^{\rm 1-loop} (k;a)$ is evaluated using eq.(\ref{regptloop}) and $\mathcal{J}(kr,\mu^2)$ is given in eq.(\ref{mybessel}). The third term contributes a constant to $\sigma_{12}^2(r,\mu^2)$. This is treated as a free parameter ($\sigma^2_{\rm iso}$) in our analysis in Sec.III (see \cite{Satpathy:2016tct} for example) but we give the PT prediction for this isotropic contribution below 
 \begin{align}  
 2b \langle \delta(\bfx)( v^{\ell}(\bfx))^2\rangle = \frac{b}{6\pi^4} \int dk dy \int_{-1}^1 & dx k^3 y^2 P_0(k) P_0(ky) G_1(k;a), \nonumber \\
 & \times \left(\frac{2G_2(ky,k,x;a) F_1(ky;a)(1+yx)}{\sqrt{1+y^2+2yx}}-\frac{xG_1(ky;a)F_2(ky,k,x;a)}{y}\right).
 \label{sigmaiso}
 \end{align} 
We can expand the 2nd line of eq.(\ref{sig12m}) as 
\begin{align}
 2b \left[\langle \delta(\bfx) (v^{\ell}(\bfx+\bfr))^2\rangle - 2 \langle \delta(\bfx)  v^{\ell}(\bfx) v^{\ell}(\bfx+\bfr) \rangle \right] = & 4b \langle \delta_1(\bfx)  v^{\ell}_1(\bfx+\bfr) v^{\ell}_2(\bfx+\bfr) \rangle + 2b \langle \delta_2(\bfx) (v^{\ell}_1(\bfx+\bfr))^2\rangle \nonumber \\ 
 & -4b  \langle \delta_1(\bfx)  v^{\ell}_1(\bfx) v^{\ell}_2(\bfx+\bfr) \rangle -4b  \langle \delta_1(\bfx)  v^{\ell}_2(\bfx) v^{\ell}_1(\bfx+\bfr) \rangle \nonumber \\ 
 &  - 4b\langle \delta_2(\bfx)  v^{\ell}_1(\bfx) v^{\ell}_1(\bfx+\bfr) \rangle.
 \end{align}
 The integrals of these terms are given below
\begin{align}
4b \langle \delta_1(\bfx)  v^{\ell}_1(\bfx+\bfr) v^{\ell}_2(\bfx+\bfr) \rangle = \frac{b}{2\pi^4} \int dk dy \int_{-1}^1 & dx k^3 y P_0(k)P_0(ky)  \frac{F_1(k;a) G_1(ky;a) G_2(ky,k,x;a)}{\sqrt{1+y^2+2yx}} \nonumber \\ & \times \left( j_0(kr)(y-2x-3x^2y) - \mathcal{J}(kr,\mu^2) y(1-x^2) \right),
\end{align}
\begin{align} 
 2b \langle \delta_2(\bfx) (v^{\ell}_1(\bfx+\bfr))^2\rangle = & -\frac{1}{16\pi^6} \int dk dy k^3 y G_1(ky;a)G_1(k;a) P_0(k) P_0(ky)
\int_{-1}^1 dx_1 dx_2  \cos{(kyrx_1 + krx_2)} \nonumber \\ &   \int_0^{2\pi} d\phi_1 d\phi_2  F_2(k,y,\bar{x};a)  \left[ \mu^2(2x_1x_2 - \bar{x}) + \bar{x}-x_1 x_2 \right],
\label{a13term}
 \end{align}
 where $\bar{x} = x_1x_2 + \sqrt{(1-x_1^2)(1-x_2^2)}\sin{\phi_1}\sin{\phi_2}$. The 4 dimensional angular integration in this expression is performed using the Monte Carlo integration algorithm Cuba \cite{Hahn:2004fe}. 
\begin{align}
-4b  \langle \delta_1(\bfx)  v^{\ell}_1(\bfx) v^{\ell}_2(\bfx+\bfr) \rangle = -\frac{b}{\pi^4} \int dk dy \int_{-1}^1 & dx k^3 y x P_0(ky) P_0(k\sqrt{1+y^2-2yx}) \mathcal{J}(kr,\mu^2) \nonumber \\ & G_1(ky;a) F_1(k\sqrt{1+y^2-2yx};a) G_2(ky,k\sqrt{1+y^2-2yx},u;a), 
\end{align}
\begin{align}
-4b  \langle \delta_1(\bfx)  v^{\ell}_2(\bfx) v^{\ell}_1(\bfx+\bfr) \rangle 
& - 4b\langle \delta_2(\bfx)  v^{\ell}_1(\bfx) v^{\ell}_1(\bfx+\bfr) \rangle \nonumber \\ &  = -\frac{b}{\pi^4} \int dk dy \int_{-1}^1 dx k^3 y P_0(k) P_0(ky) \mathcal{J}(kr,\mu^2) \nonumber \\ & \times G_1(k;a)\left(G_2(ky,k,x;a)F_1(ky;a)\frac{y(1+yx)}{1+y^2+2yx} - x G_1(ky;a)F_2(ky,k,x;a)\right). 
\end{align}
Finally, the last term in eq.(\ref{sig12m}) evaluates to 
\begin{equation}
2b^2 \left[ \langle \delta(\bfx) \delta(\bfx + \bfr)  (v^{\ell}(\bfx))^2 \rangle - \langle \delta(\bfx) \delta(\bfx + \bfr)  v^{\ell}(\bfx)  v^{\ell}(\bfx + \bfr) \rangle \right]  = b^2\xi^r(r) \sigma_{12,{\rm lin}}^2(r,\mu^2) + \frac{1}{2}v_{12,{\rm lin}}^2(r) \mu^2, 
\label{sig12last}
\end{equation} 
where $\sigma_{12,{\rm lin}}^2(r,\mu^2) $ and $v_{12,{\rm lin}}^2(r)$ are the linear predictions for the velocity dispersion eq.(\ref{lins12}) and mean infall velocity eq.(\ref{linv12}).  At leading order the first term in eq.(\ref{sig12last}) cancels with the 2nd term on the LHS of eq.(\ref{sig12m}) and so we omit in our calculations and simply include the linear mean infall velocity term.
\newline
\newline
For the calculations in the next section we have set $b=1$ and so only dark matter particles are considered. Perturbation theory predicts a constant contribution to the velocity dispersion, $\sigma^2_{\rm iso}$ (eq.(\ref{sigmaiso})) given in units of (Mpc/$h$)${}^2$. As mentioned, this is treated as a free parameter allowing us to describe deviations to the predicted scale dependance on small scales where non-linear fingers of god effects are strong and unable to be treated perturbatively.  
\newpage
\section{Results}
\begin{figure}[H]
  \captionsetup[subfigure]{labelformat=empty}
  \centering
  \subfloat[]{\includegraphics[width=10.3cm, height=8.5cm]{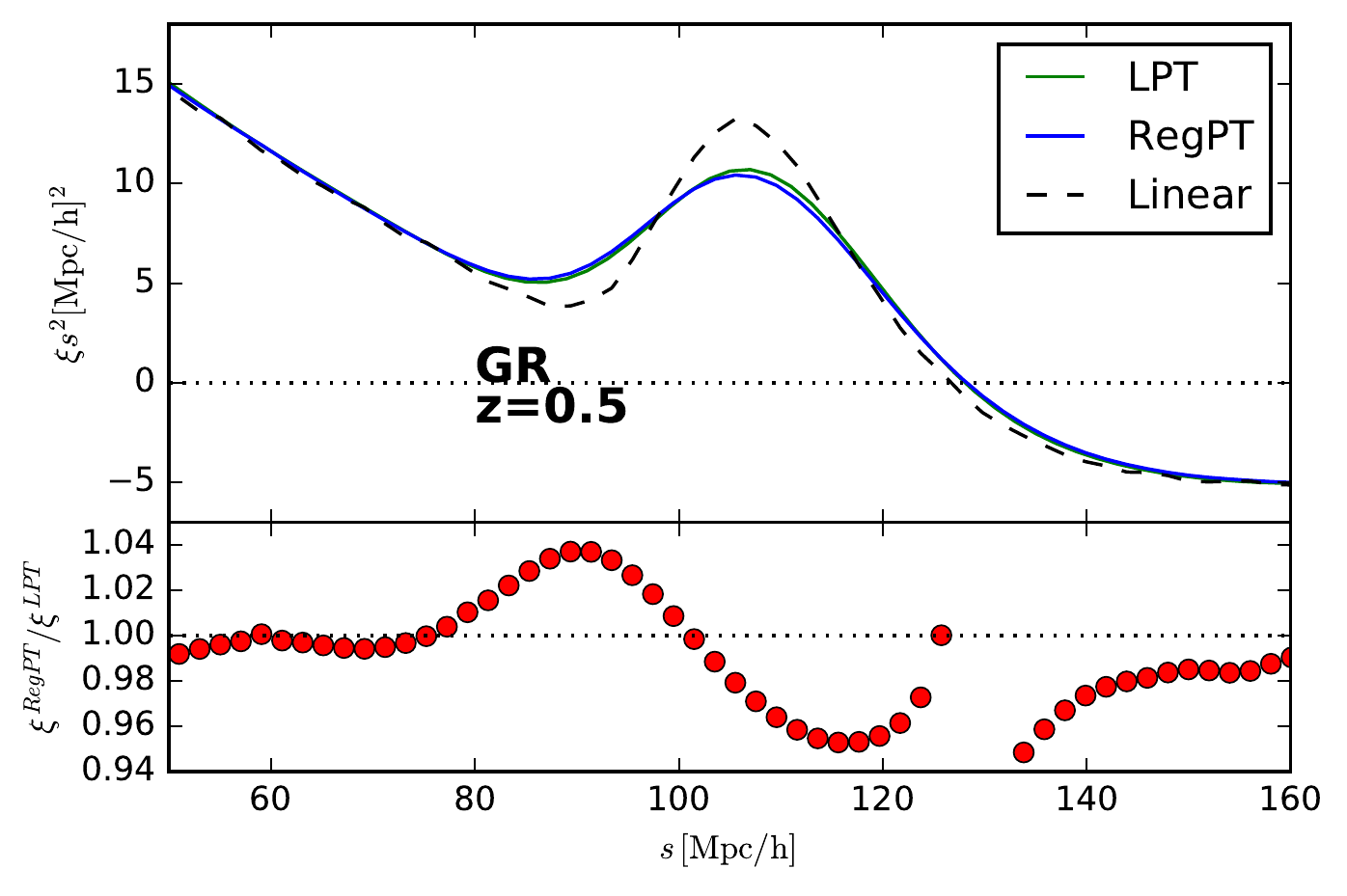}}
  \caption[CONVERGENCE ]{Comparison of real space predictions for the correlation function using LPT (green), FT of RegPT (blue) and Linear (black,dashed). The bottom panel shows the fractional difference between the LPT and FT of RegPT. The reader should keep in mind that there is a 0-crossing at $r=130$Mpc/$h$ causing large fractional differences.}
\label{xir1}
\end{figure}

\begin{figure}[H]
  \captionsetup[subfigure]{labelformat=empty}
  \centering
  \subfloat[]{\includegraphics[width=8.3cm, height=8.5cm]{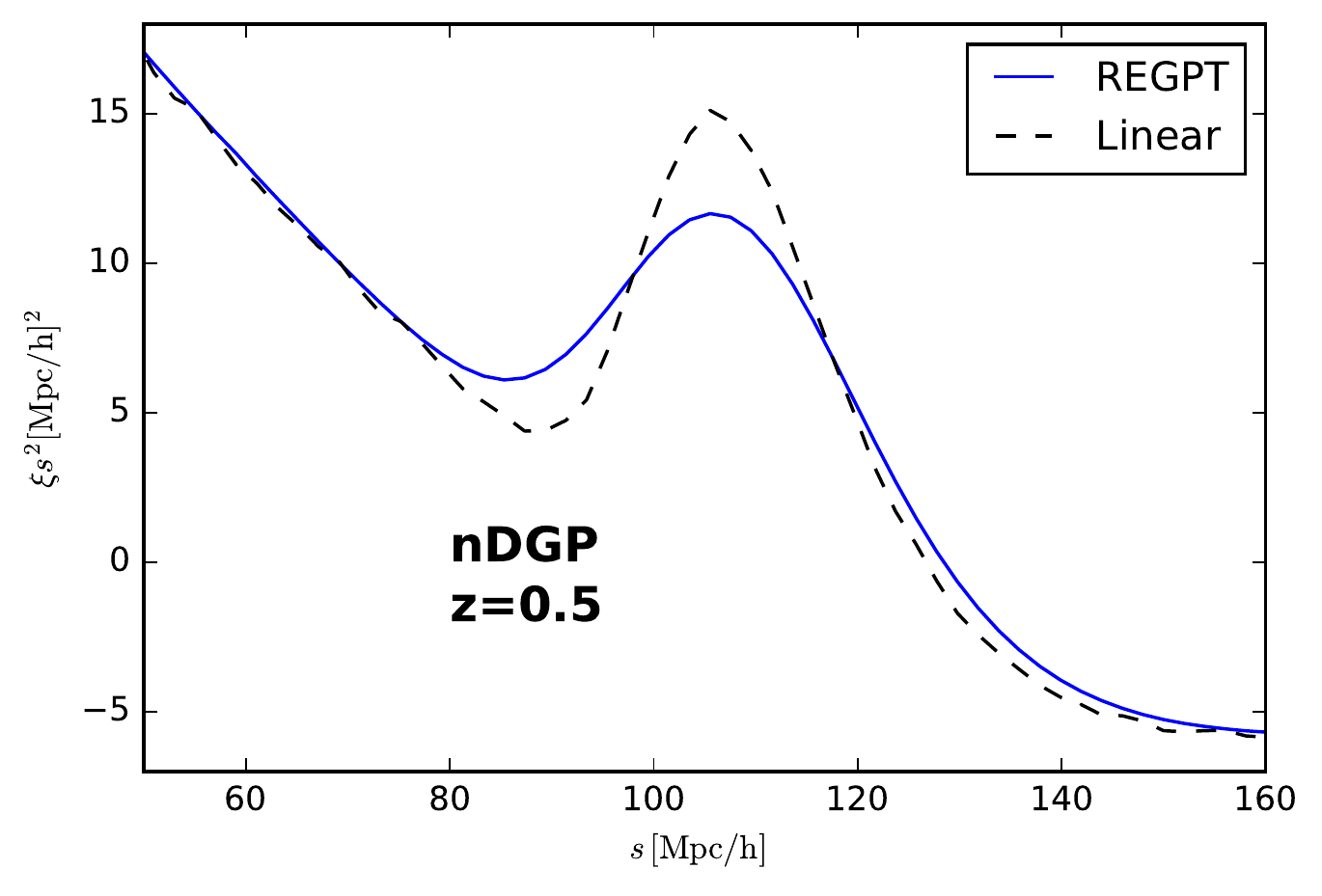}} \quad
    \subfloat[]{\includegraphics[width=8.3cm, height=8.5cm]{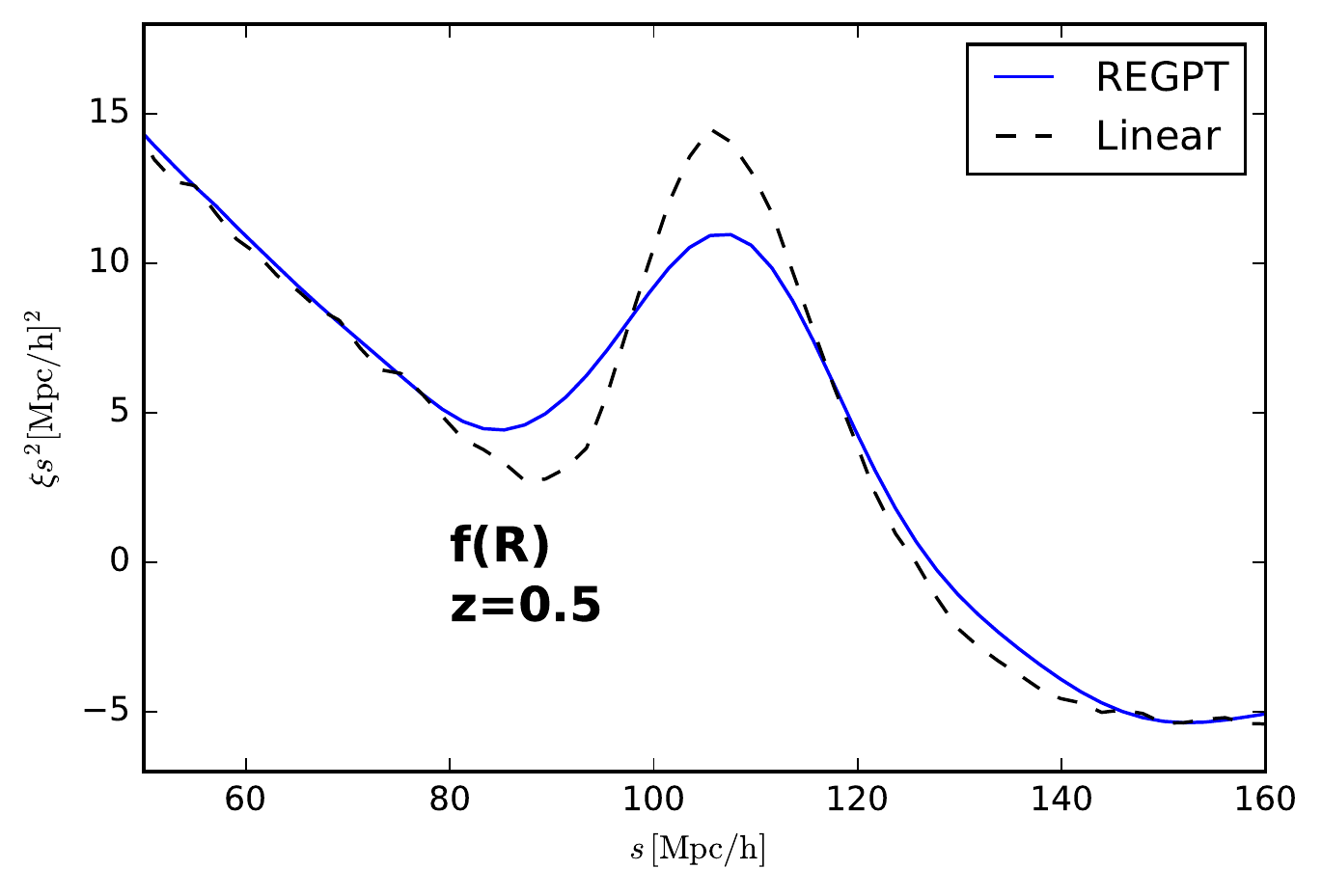}} 
  \caption[CONVERGENCE ]{Comparison of real space predictions for the correlation function using the FT of the 1-loop power spectrum RegPT (blue) and FT of linear power spectrum (black,dashed) for nDGP (left) and $f(R)$ (right). }
\label{xir2}
\end{figure}

\begin{figure}[H]
  \captionsetup[subfigure]{labelformat=empty}
  \centering
  \subfloat[]{\includegraphics[width=8.3cm, height=8.5cm]{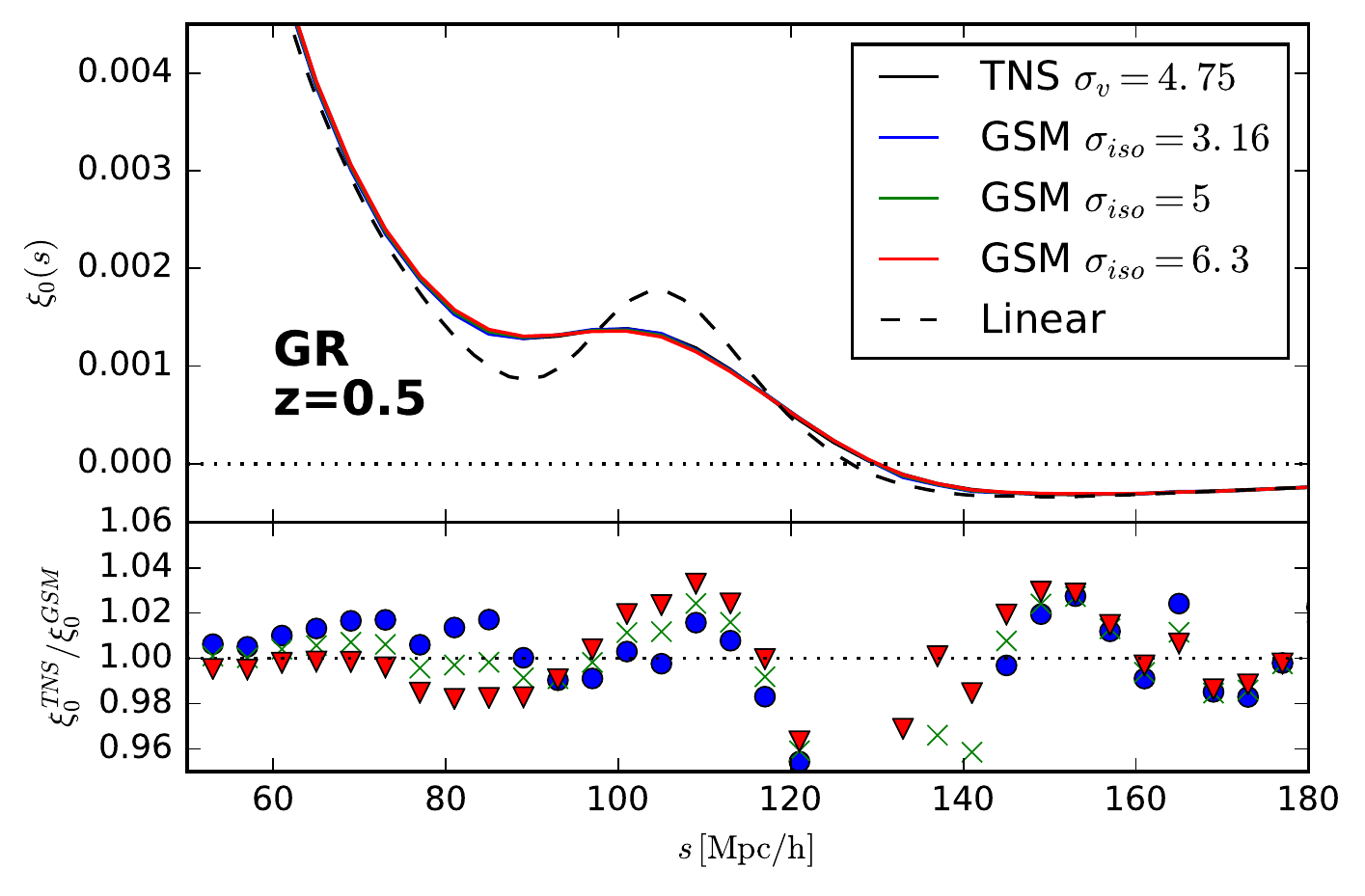}} \quad
    \subfloat[]{\includegraphics[width=8.3cm, height=8.5cm]{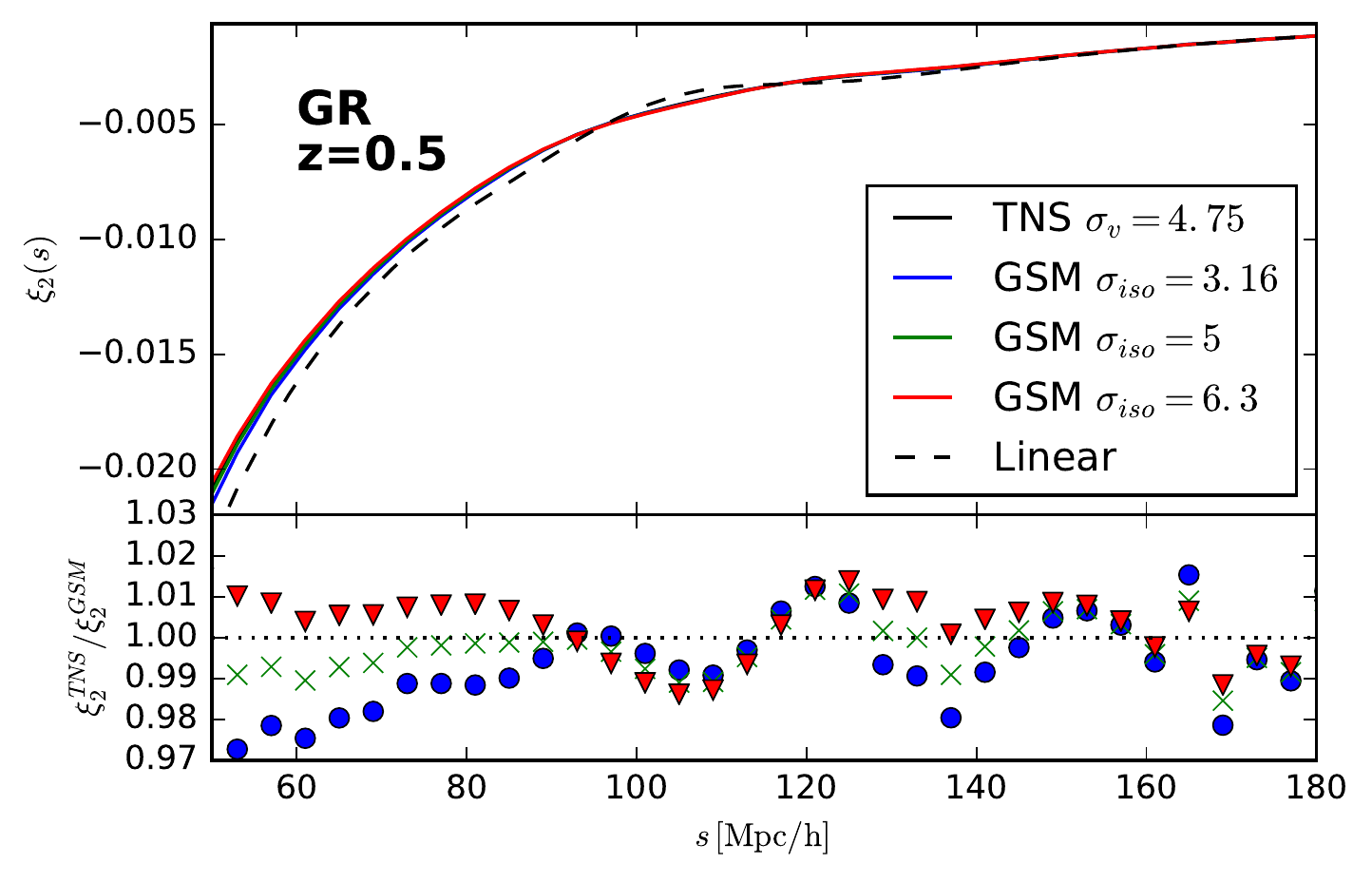}} 
  \caption[CONVERGENCE ]{Comparison of the redshift space predictions for the correlation function using the FT of the TNS power spectrum (black solid) with $\sigma_v=4.75$ Mpc/$h$ and the GSM for three values of $\sigma_{\rm iso}$ (in units of Mpc/$h$) for GR. The PT prediction for $\sigma_{\rm iso}$ is given by the blue curve. The LSM prediction is shown as a dashed black curve. The left plot shows the monopole while the right plot shows quadrupole. The bottom panels shown the fractional difference between the TNS transform and the GSM. Keep in mind the zero crossing indicated by the dotted line in the top panel of the monopole giving large fractional differences. }
\label{xirs1}
\end{figure}

\begin{figure}[H]
  \captionsetup[subfigure]{labelformat=empty}
  \centering
  \subfloat[]{\includegraphics[width=8.3cm, height=8.5cm]{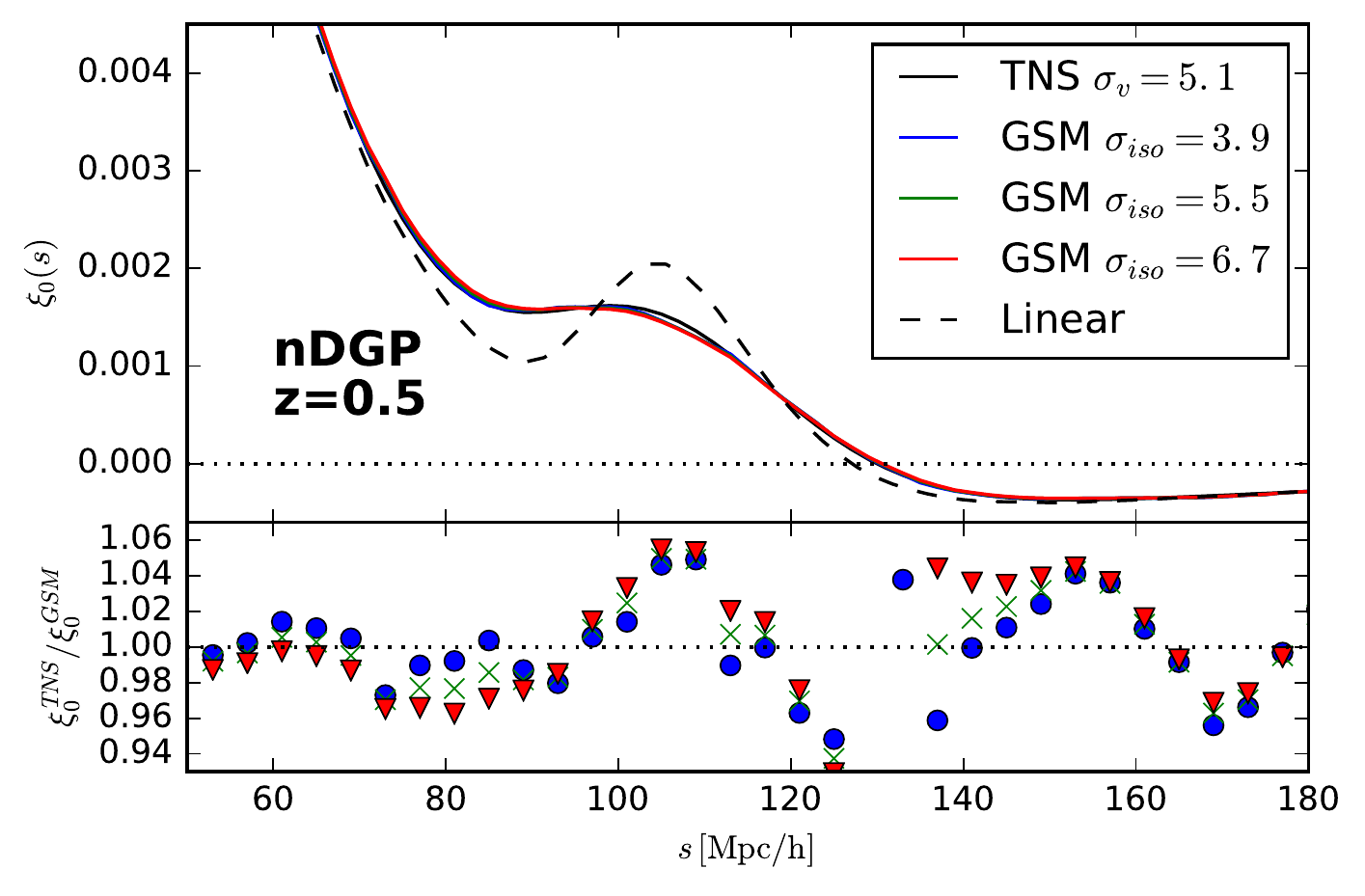}} \quad
    \subfloat[]{\includegraphics[width=8.3cm, height=8.5cm]{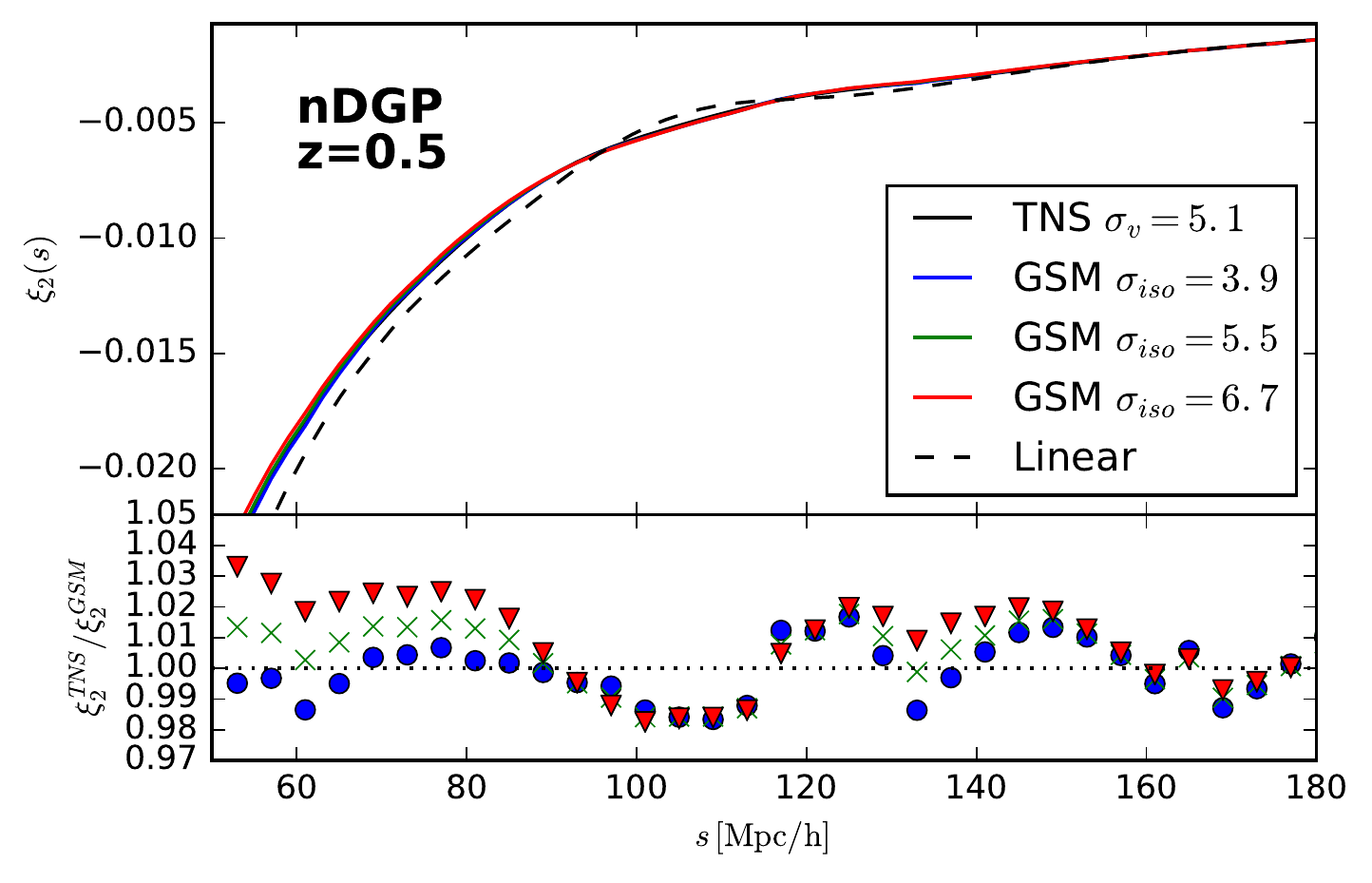}} 
  \caption[CONVERGENCE ]{Same as Fig.\ref{xirs1} but for the nDGP model of gravity with $\Omega_{rc} = 0.438$. The TNS transform uses $\sigma_v=5.1$ Mpc/$h$.}
\label{xirs2}
\end{figure}

\begin{figure}[H]
  \captionsetup[subfigure]{labelformat=empty}
  \centering
  \subfloat[]{\includegraphics[width=8.3cm, height=8.5cm]{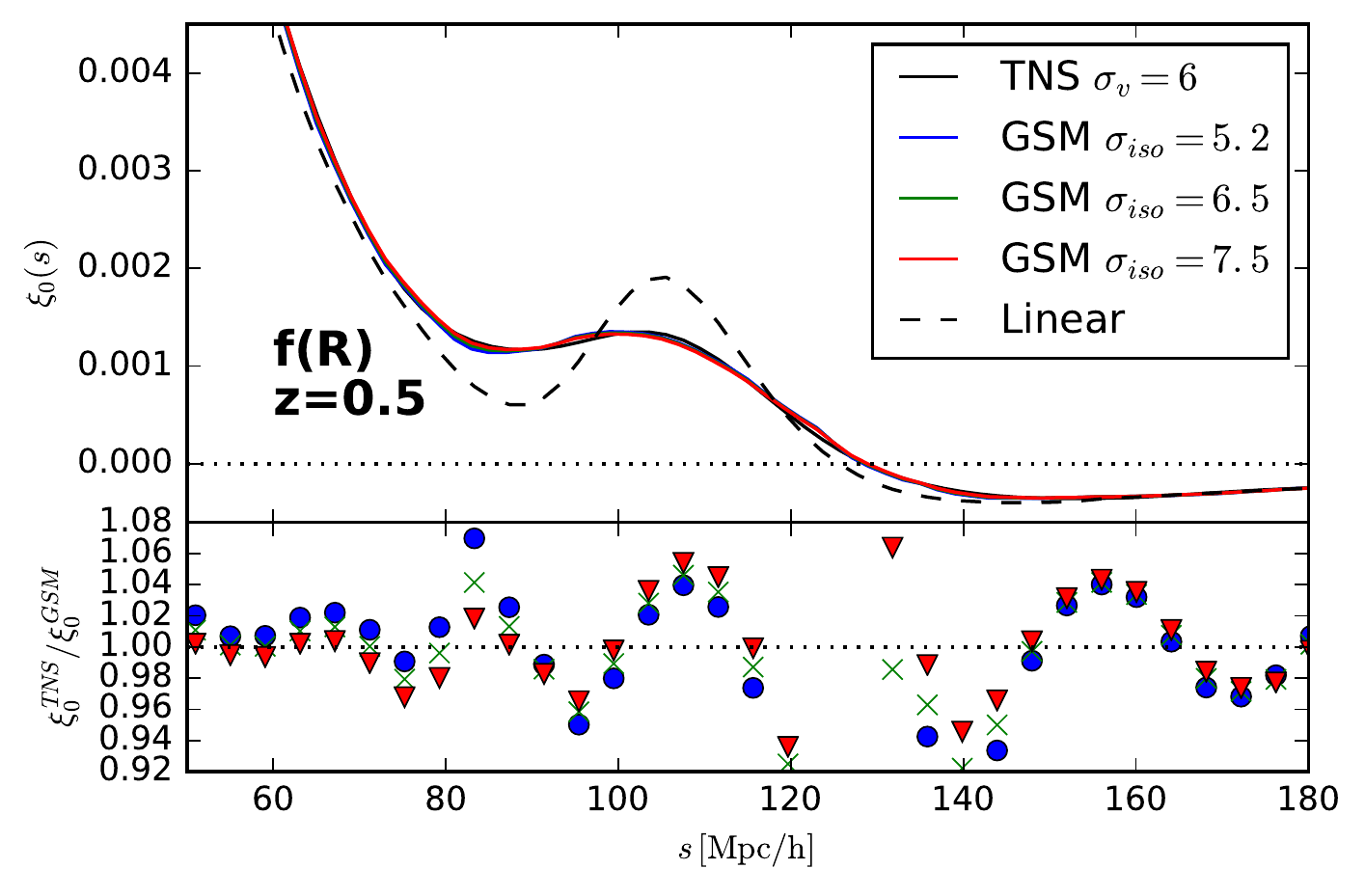}} \quad
    \subfloat[]{\includegraphics[width=8.3cm, height=8.5cm]{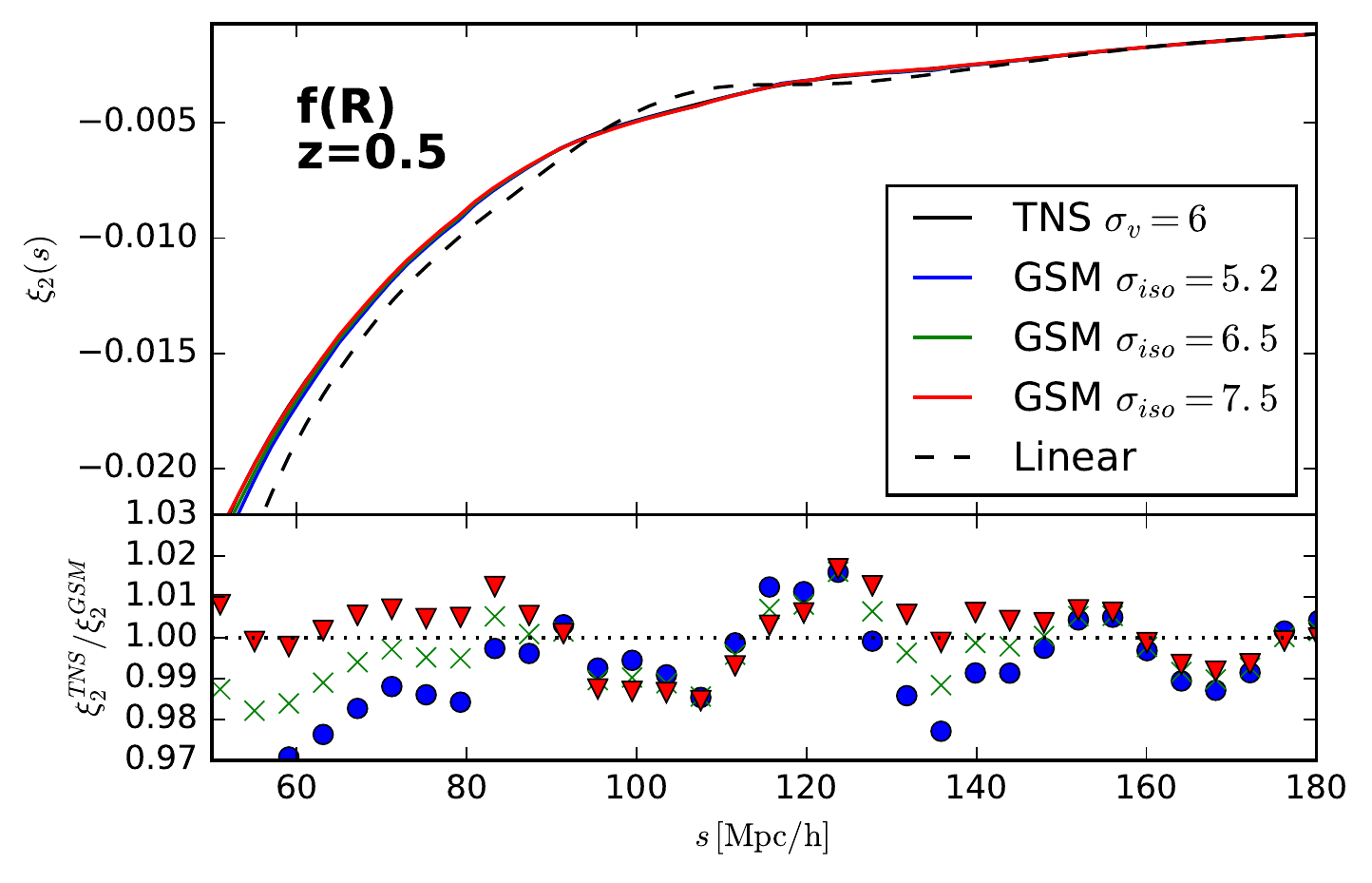}} 
  \caption[CONVERGENCE ]{ Same as Fig.\ref{xirs1} but for the Hu-Sawicki model of $f(R)$ gravity with $|f_{R0}|=10^{-4}$. The TNS transform uses $\sigma_v=6$ Mpc/$h$.}
\label{xirs3}
\end{figure}

\begin{figure}[H]
  \captionsetup[subfigure]{labelformat=empty}
  \centering
  \subfloat[]{\includegraphics[width=8.3cm, height=8.5cm]{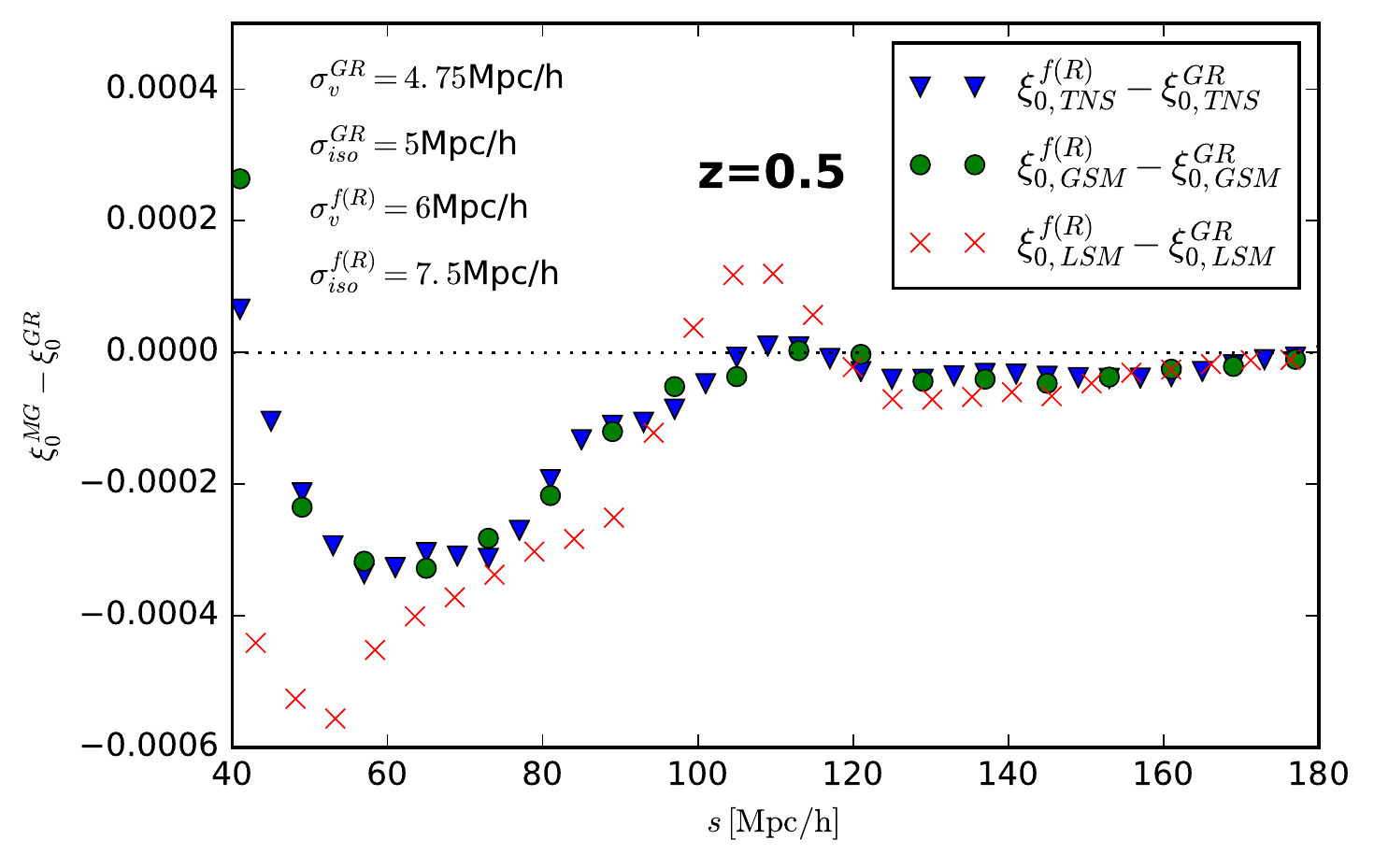}} \quad
    \subfloat[]{\includegraphics[width=8.3cm, height=8.5cm]{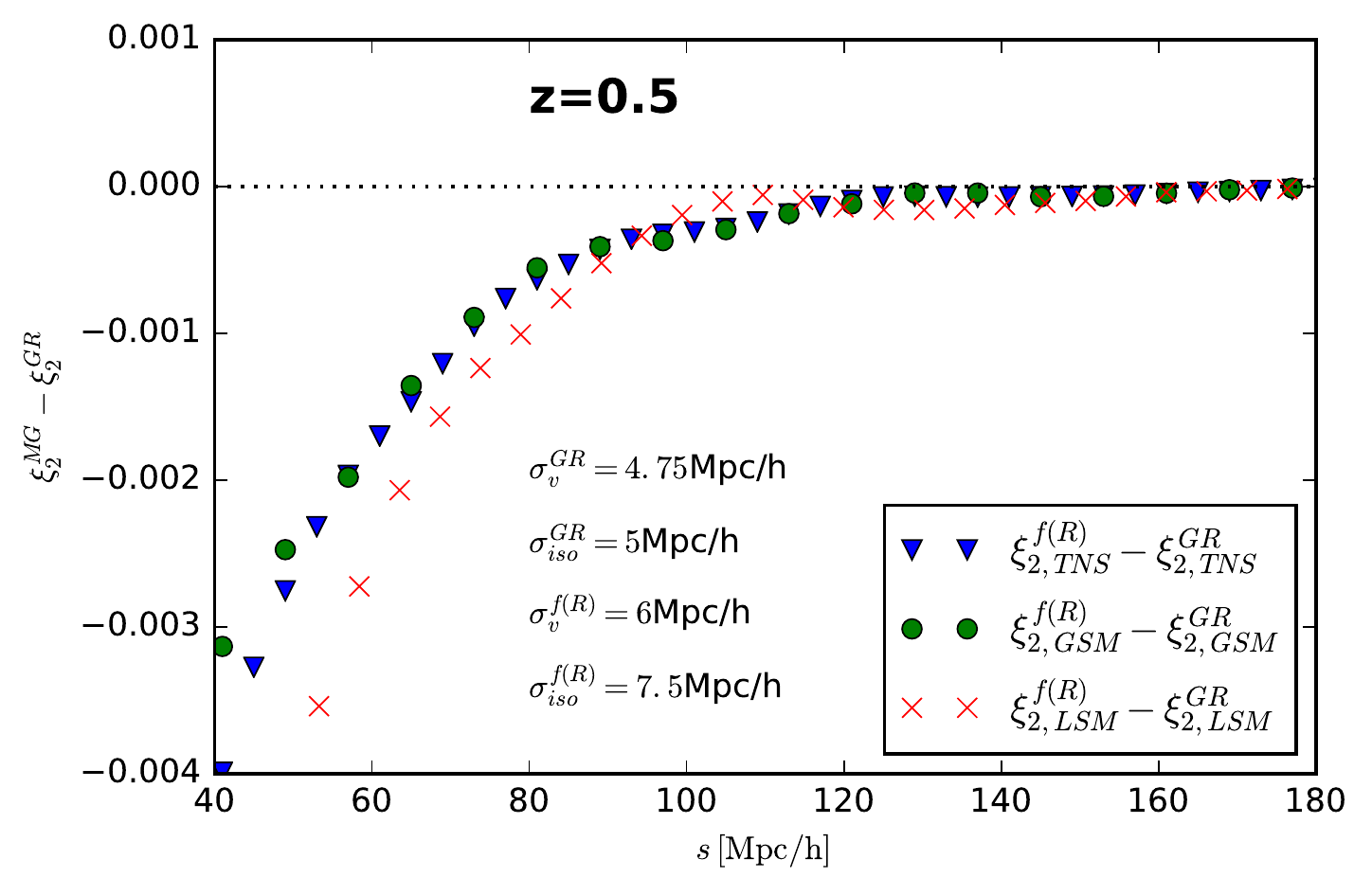}} 
  \caption[CONVERGENCE ]{The difference in the GSM (green), FT of the TNS best fit (blue) and LSM(red) multipoles between $f(R)$ and GR at $z=0.5$. The monopole difference is shown on the left and the quadrupole difference is shown on the right. }
\label{frvgr}
\end{figure}

\begin{figure}[H]
  \captionsetup[subfigure]{labelformat=empty}
  \centering
  \subfloat[]{\includegraphics[width=8.3cm, height=8.5cm]{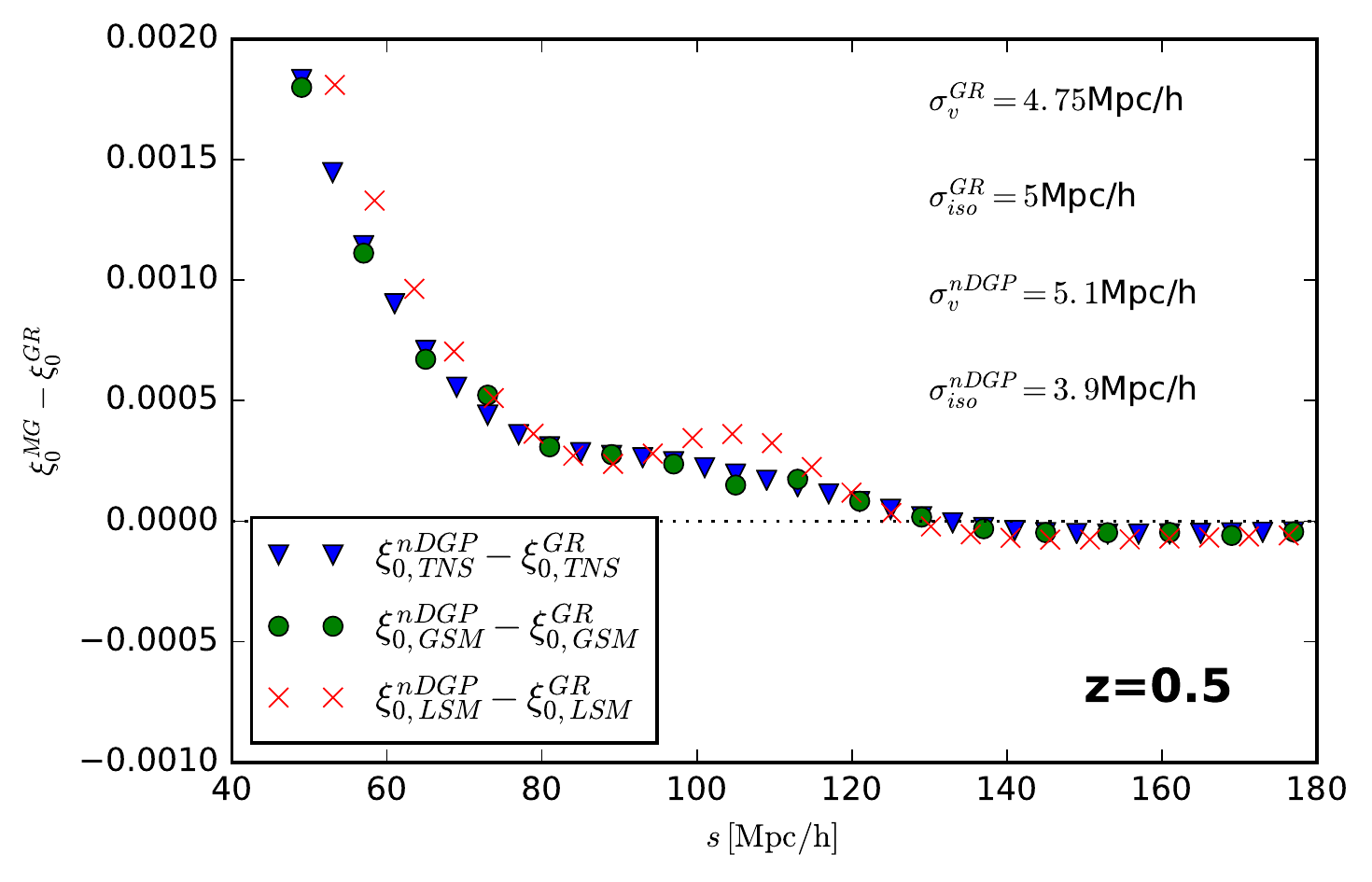}} \quad
    \subfloat[]{\includegraphics[width=8.3cm, height=8.5cm]{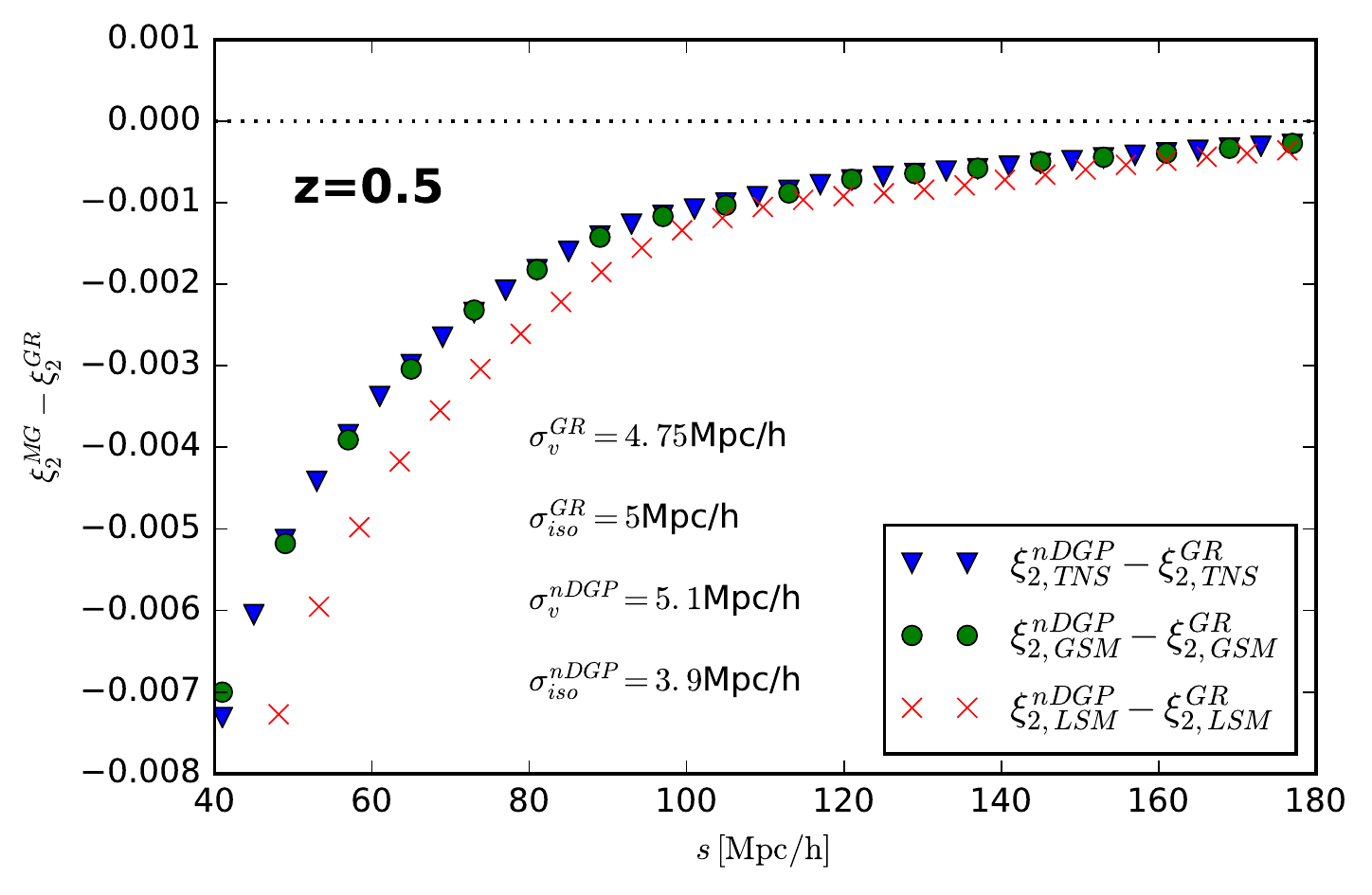}} 
  \caption[CONVERGENCE ]{The difference in the GSM (green), FT of the TNS best fit (blue) and LSM(red) multipoles between nDGP and GR at $z=0.5$. The monopole difference is shown on the left and the quadrupole difference is shown on the right. }
\label{dgpvgr}
\end{figure}

In this section we will present predictions using eq.(\ref{redshiftcor}) for three models of gravity, namely the Vainshtein screened normal branch of DGP gravity (nDGP) \cite{Dvali:2000hr}, the Chameleon screened Hu-Sawicki form of $f(R)$ gravity \cite{Hu:2007nk} and GR. We will compare these results with the FT of eq.(\ref{redshiftps}) which is fit to N-body simulations. This is done for dark matter only and no tracer bias is included.  The Fourier space comparisons for nDGP can be found in Appendix A while for GR and $f(R)$ we use the best fit $\sigma_v$ found in \cite{Taruya:2014faa}.
\newline
\newline
Our background cosmology is taken from WMAP9~\citep{Hinshaw:2012aka}: $\Omega_m = 0.281$, $h=0.697$, and $n_s=0.971$. The box width is $1024 \mbox{Mpc}/h$ with $1024^3$ dark matter particles used and a starting redshift of $49$. The linear theory power spectrum normalisation was set to be $\sigma_8=0.844$. The nDGP simulation uses $\Omega_{rc}=1/4r_c^2H_0^2=0.438$ while the $f(R)$ simulation uses $|f_{R0}|=10^{-4}$. We consider the redshift of $z=0.5$ where SPT benefits from a good range of validity while still being relevant for upcoming surveys such as Euclid \footnote{\url{www.euclid-ec.org}} and DESI  \footnote{\url{http://desi.lbl.gov/}}. We start with a comparison of linear and non-linear predictions for the real space correlation function followed by comparing different predictions for the non-linear redshift space correlation function predictions: the GSM using RegPT and the FT of TNS multipoles. 
\newline
\newline
We compare the FT of the RegPT 1-loop expressions with the LPT model of \cite{Matsubara:2008wx}. This model has been tested against N-body simulations in the GR case and has shown to be percent level accurate at scales of $r\geq25$ Mpc/$h$ \cite{Reid:2011ar}. It has been employed in spectroscopic survey analyses with BOSS \cite{Reid:2012sw}. Although we only do this for GR, it gives us a handle on the accuracy of our FT approach to the multipoles. The transform of the RegPT power spectrum was compared to N-body results for GR and $f(R)$ in \cite{Taruya:2014faa} showing  good agreement above and around the BAO scale.
\newline
\newline
Fig.\ref{xir1} shows the real space correlation function as predicted by eq.(\ref{pregab}) and the LPT prediction of  \cite{Matsubara:2008wx} for dark matter. The FT of the linear power spectrum is also shown as the linear prediction. We see both RegPT and LPT give a smoothing of the BAO bump - a well known non-linear effect - and that they agree on small and large scales at the percent level while around the BAO bump they show up to a $4\%$ difference with the RegPT treatment showing slightly more damping around this scale. For completeness we  also show the RegPT predictions against the linear predictions for the other models of gravity considered (Fig.\ref{xir2}). We notice that the non-linear RegPT predictions for these models show more damping of the BAO bump when compared to the GR case, an expected effect of enhanced structure growth as well as enhanced 2nd and 3rd order non-linearities. 
\newline
\newline
Moving to redshift space, we will use the FT of the best fit multipoles shown on the left of Fig.\ref{pab1} in the Appendix for nDGP and Table. II of \cite{Taruya:2014faa} for $f(R)$ and GR. Because of the TNS's extra degree of freedom ($\sigma_v$), the model should have an advantage in goodness of fit when compared to the GSM, which can be completely determined by SPT. In general the correlation function needs to be measured many times from N-body simulations and averaged because of the small imprint of the acoustic features which can be greatly hidden by scatter. MG simulations are more computationally expensive than GR ones and so only a few are available. Thus, a clean configuration space measurement in MG theories is not readily available. This makes the TNS transform a good and practical benchmark to compare the GSM predictions to in the absence of averaged N-body correlation function measurements. The configuration space multipoles are given by \cite{Cole:1993kh,Hamilton:1997zq} 
\begin{equation}
\xi_\ell^{(S)}(s)= \frac{i^\ell}{2\pi^2} \int dk k^2 P_\ell^{(S)}(k)j_\ell(ks),
\label{multipolesR}
\end{equation}
where $j_\ell$ is the $\ell^{\rm th}$ order spherical Bessel function and $P_\ell^{(S)}$ is given by
\begin{equation}
P_\ell^{(S)}(k)=\frac{2\ell+1}{2}\int^1_{-1}d\mu P_{\rm TNS}^{(S)}(k,\mu)\mathcal{P}_\ell(\mu),
\label{multipolesF}
\end{equation}
\newpage
\noindent where $\mathcal{P}_l(\mu)$ denote the Legendre polynomials. Again we will only consider the first two multipoles, $\ell = 0,2$. The top panel of Fig.\ref{xirs1} shows the monopole (left) and quadrupole (right) predictions for the redshift space correlation function within GR. We have plotted the TNS transform with $\sigma_v = 4.75$Mpc/$h$ in black against the GSM predictions for three different values of the parameter $\sigma_{\rm iso}$ defined in Sec.II C. The blue curve is the GSM prediction where $\sigma_{\rm iso}$ takes the PT predicted value. The predictions look very reasonable with significant smearing of the BAO due to non-linear effects, mostly seen in the monopole. 
\newline
\newline
The bottom panels of Fig.\ref{xirs1} show the fractional differences between the TNS transform and the GSM predictions. Fractional differences go up to $4\%$ in the monopole around the BAO scale and slightly less for the quadrupole, with slightly more damping of the BAO bump by the GSM predictions. We find that around this scale the PT prediction (eq.(\ref{sigmaiso})) for the isotropic contribution to the velocity dispersion does well for the monopole, whereas for the quadrupole the higher valued green curve ($\sigma_{\rm iso} = 5 $Mpc/$h$) does better, a value consistent with the TNS best fit velocity dispersion.  
\newline
\newline
Similar results are found for the nDGP model of gravity, shown in Fig.\ref{xirs2}. The deviations of the GSM predictions from the TNS transform are only slightly larger than in the GR case, going up to $6\%$ in the monopole at the BAO scale. The PT prediction for $\sigma_{\rm iso}$ ($\sigma_{\rm iso} = 3.9 $Mpc/$h$) does the best over both multipoles at smaller scales with the green ($\sigma_{\rm iso} = 5.5 $Mpc/$h$) doing a bit better around the BAO bump. Both these values are consistent with the TNS best-fit value. 
\newline
\newline
The $f(R)$ predictions are shown in Fig.\ref{xirs3}. In this case the monopole's fractional differences are significantly larger with up to $8\%$ more damping in the GSM model. The quadrupole differences remain $\leq 3\%$ around the BAO scale. In this case the PT predicted value for $\sigma_{\rm iso}$ ($5.2 $Mpc/$h$) seems to underestimate the value with  $\sigma_{\rm iso} = 7.5 $Mpc/$h$ being more consistent with the TNS transform. This being said, to really tell which treatment performs better we wait for comparisons with simulation data. As mentioned earlier, many realisations are needed to get a converged measurement of the correlation function. This can be done for GR but for MG theories simulations are expensive computationally. By using COmoving Lagrangian Acceleration (COLA) approaches such as those described in \cite{Winther:2017jof}, this problem becomes tractable and we leave this to a future work. 
\newline
\newline
Fig.\ref{frvgr} and Fig.\ref{dgpvgr} show the differences between the modified gravity predictions and the GR ones for both theoretical predictions for the correlation function as well as the linear prediction. We see that in both the FT of TNS and GSM the effect of modifying gravity is very similar indicating both approaches give comparable signals of deviations from GR.  In the monopole, around the acoustic bump, both non-linear approaches reduce the MG-Signal with a larger difference seen in linear modelling. The LSM also shows larger differences at scales below the BAO in the quadrupole. One other feature is that $f(R)$ gravity shows a suppression compared to GR around  $40$ Mpc$/h \leq s \leq100$ Mpc$/h $  while nDGP shows an enhancement over GR for the monopole. 
%% Comment on this
\newpage
%%%%%%%%%%%%%%%%%%%%%%%%%%%%%%%%%
\section{Summary}
This work has extended the code described in \cite{Bose:2016qun} to calculate the non-linear redshift space correlation function as modelled by \cite{Reid:2011ar} for a general class of gravity and dark energy models. We have also extended the code to calculate the non-linear redshift space correlation function as described by the TNS model using the RegPT treatment as done in \cite{Taruya:2014faa}. To make comparisons between the two predictions the TNS power spectrum monopole and quadrupole were first compared to N-body data in order to obtain the best fit $\sigma_v$ (See Fig.\ref{pab1} and Table.II of \cite{Taruya:2014faa}). This required finding a realm of validity for the SPT predictions which was found by comparing the real space power spectra (See Appendix A). We then found fair agreement between these two treatments to within $4\%$ for GR and nDGP with $\Omega_{rc} = 0.438$ and up to a $8\%$ deviation in the treatments for the chameleon screened $f(R)$ model with $|f_{R0}|=10^{-4}$ around the BAO scale (Fig.\ref{xirs1}, Fig.\ref{xirs2} and Fig.\ref{xirs3}). We have also compared the LPT correlation function \cite{Matsubara:2008wx} in real space with that obtained using a FT of the RegPT 1-loop spectrum (Fig.\ref{xir1}). The RegPT treatment gives up to $4\%$ more damping around the BAO scale. Recently a LPT prediction for MG models has been developed \cite{Aviles:2017aor} allowing the extension of such comparisons. 
\newline
\newline
We observe large damping in the GSM and FT of TNS treatments over the linear predictions with more damping observed in the $f(R)$ and nDGP cases. This is due to gravity being boosted by additional non-linearities encoded in the extra $\gamma$ functions for these theories. The difference between the GSM and FT of TNS predictions comes from their treatment of the RSD. While the GSM is completely perturbative in making the non-linear mapping to redshift space within configuration space, the TNS is partly phenomenological and further, a resummation technique such as RegPT is needed to make the transform to configuration space. Despite its added degree of freedom, $\sigma_v$, it is unclear how best to treat small scale SPT divergences and further how robust and consistent the methods on the market are (examples of such treatments include RegPT \cite{Taruya:2012ut}, renormalised perturbation theory \cite{Crocce:2005xy,Crocce:2007dt} and EFToLSS prescriptions \cite{Senatore:2014via,Vlah:2015sea}). This issue has yet to be investigated thoroughly. In light of this, one cannot say with certainty which approach to the redshift space correlation function will perform better when matching simulation or observational data. This will be the focus of a future work.  
\newline
\newline
To give the GSM model extra freedom, we promote the isotropic velocity dispersion contribution to the GSM's pairwise dispersion $\sigma_{12}^2$  as a free parameter $\sigma_{\rm iso}$, which is physically equivalent to TNS's $\sigma_v$ parameter. By doing this we can enhance the PT prediction, given in eq.(\ref{sigmaiso}), and better match the TNS on small scales. We find that the PT prediction for $\sigma_{\rm iso} = 3.9$ Mpc/$h$ does well for the nDGP model and we are able to match the FT of TNS prediction at scales $s \leq 100$Mpc$/h$ to within $2\%$ (Fig.\ref{xirs2}).  For $f(R)$ and GR we find the PT prediction underestimates the small scale velocity dispersion, and we find the larger values of $\sigma_{\rm iso}^{\rm GR} = 5 (3.16)$ Mpc/$h$ and $\sigma_{\rm iso}^{f(R)} = 7.5(5.2)$ Mpc/$h$ (PT prediction in brackets) better match the FT of TNS at smaller scales, specifically in the quadrupole prediction (Fig.\ref{xirs1} and Fig.\ref{xirs3}). Around the scales $100 $ Mpc/$h \leq s \leq 180 $ Mpc/$h$ $\sigma_{\rm iso}$ has a marginal effect. The preferred values of $\sigma_{\rm iso}$  in the modified gravity theories both differ by around $30 \%$ when compared with the best fit values of $\sigma_v$ of the TNS model. The GR value of $\sigma_{\rm iso}$ is within $\sim 5\%$ of its TNS equivalent. In summary, we find that both approaches model the RSD consistently in the range $50\mbox{Mpc}/h \leq s \leq 180 \mbox{Mpc}/h$ with  the GSM requiring the promotion of $\sigma_{12}^2$ to a free parameter to be consistent with the TNS approach, particularly for the quadrupole. 
\newline
\newline
Using the best fit values for $\sigma_{\rm iso}$ we find that the differences between GR and MG-GSM predictions for the correlation function multipoles accurately follow those using the FT of TNS indicating that both approaches to modelling the RSD consistently treat modifications to gravity, with neither giving an enhanced MG signal over the other (Fig.\ref{frvgr} and Fig.\ref{dgpvgr}). The non-linear differences follow the LSM differences in all cases with the LSM generally picking up larger deviations from GR consistently in both multipoles. This may be because of MG's enhanced non-linear gravitation which suppresses enhancements in the multipoles.
\newline
\newline
The survey comparisons done in \cite{Reid:2012sw} imply the GSM treatment overdamps the BAO wiggle in redshift space. This suggests a preference of the RegPT treatment to the real and redshift space correlation function although marginally. Again, we wait for the availability of simulation data to make this conclusion. In any case, the ability to compute the redshift space correlation function for generalised models should prove to be very useful when performing statistical analyses on survey data and obtaining gravitational parameter constraints. The importance of correctly modelling gravity has been investigated in a number of works  \cite{Taruya:2013my,Barreira:2016mg,Bose:2017myh} and has been shown to be of growing importance as we enter the era of stage IV surveys. Using the pipeline described here we can perform consistent analyses of the high quality upcoming data from surveys such as  the Dark Energy Spectroscopic Instrument (DESI) \footnote{\url{http://desi.lbl.gov/}} and the ESA/Euclid survey\footnote{\url{www.euclid-ec.org}}. 
\newline
\newline
Further, by moving to smaller scales and using a fuller shape of the correlation function we expect any deviations from GR to become less able to hide in nuisance degrees of freedom such as $\sigma_v$, $\sigma_{\rm iso}$ or tracer bias. Fig.\ref{frvgr} and Fig.\ref{dgpvgr} show the difference between the MG and GR predictions for the correlation function. We see that at the BAO scale down to the scales valid for the GSM treatment, we have a significant MG signal.  By pushing into these scales we enter regions as yet unused for constraining models beyond GR \cite{Baker:2014zba}. This work primes the consistent probing of parameter space in this regime by using currently available spectroscopic data such as BOSS and further the possibility of using a combination of configuration and Fourier space measurements which will be very useful in overcoming systematics.  
\newline
\newline
Finally we comment on the preparation of the code for such statistical analyses. Currently optimisation needs to be made in the computation of eq.(\ref{a13term}) which on top of the 2 spatial integrals and 1 angular for the multipoles, 4 additional angular integrals need to be performed. The current method is to use an Monte Carlo integration technique to evaluate the integral which is slow when looking to achieve the desired accuracy. For statistical analyses of data a lower time cost is essential.  Further, for scale dependent models of gravity, the perturbation kernels need to be initialised many times which also incurs a significant time cost. These issues have been relieved  to some extent through parallelisation. One can also perform an interpolation technique in gravitational parameter space as done for the BOSS analysis in \cite{Song:2015oza} which reduces the number of model computations significantly. We aim to optimise the computation of eq.(\ref{a13term}) and make use of the code to perform analyses of MG models with currently available data in a future work.
%comment on biasing and GSM extensions 
%%%%%%%%%%%%%%%%%%%%%%%%%%%%%%%%
\section*{Acknowledgments}
\noindent The authors would like to thank Yuting Wang for useful discussions. We would like to thank Gong-bo Zhao and Wojciech Hellwing for supplying us with the N-body data used in Appendix A. BB is supported by the University of Portsmouth. KK is supported by the European Research Council through 646702 (CosTesGrav). KK is also supported by the UK Science and Technologies Facilities Council grants ST/N000668/1.
\newpage
\appendix
\section{Fourier Space Comparisons : nDGP} 
\begin{figure}[H]
  \captionsetup[subfigure]{labelformat=empty}
  \centering
  \subfloat[]{\includegraphics[width=8.3cm, height=8.5cm]{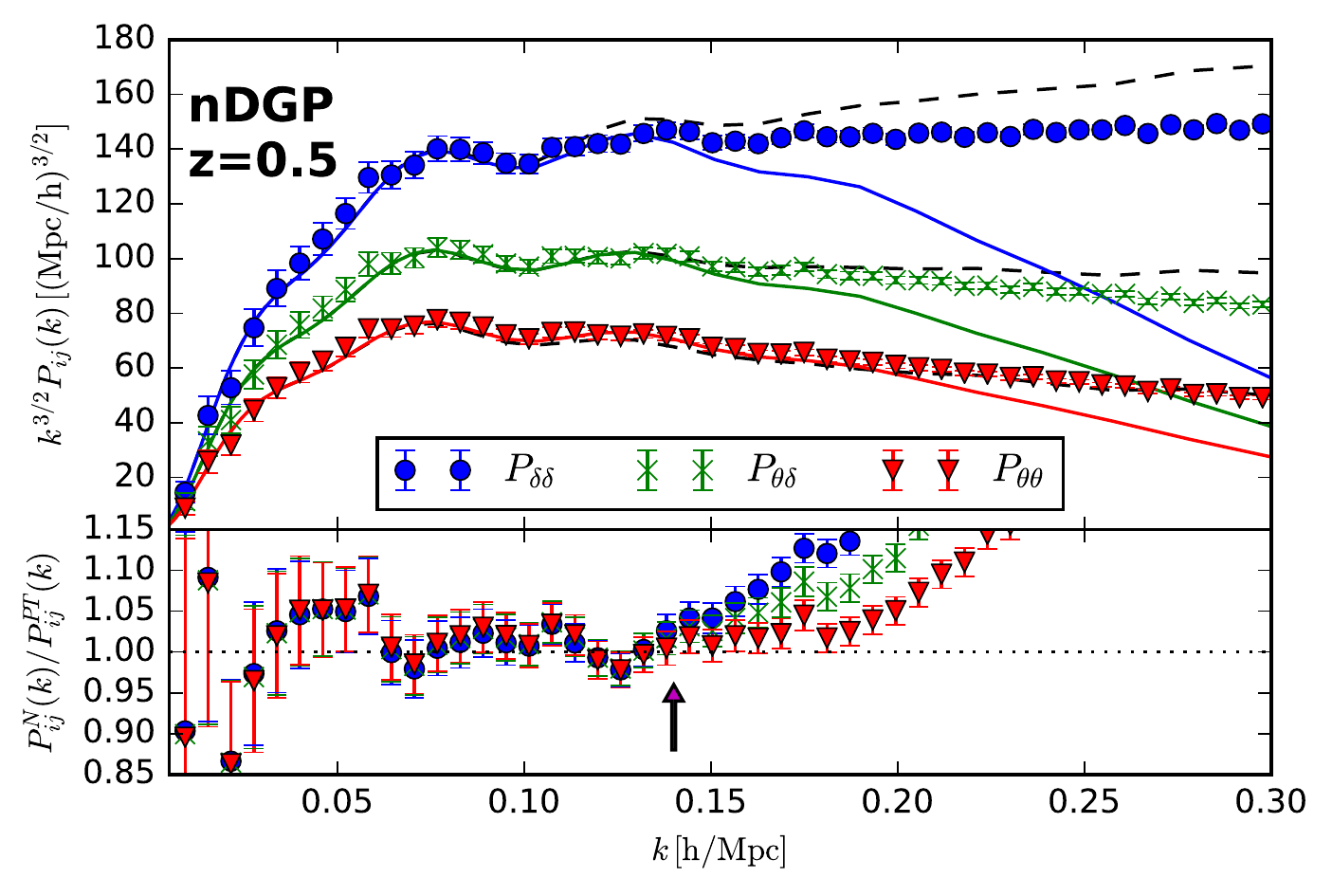}} \quad
  \subfloat[]{\includegraphics[width=8.3cm, height=8.5cm]{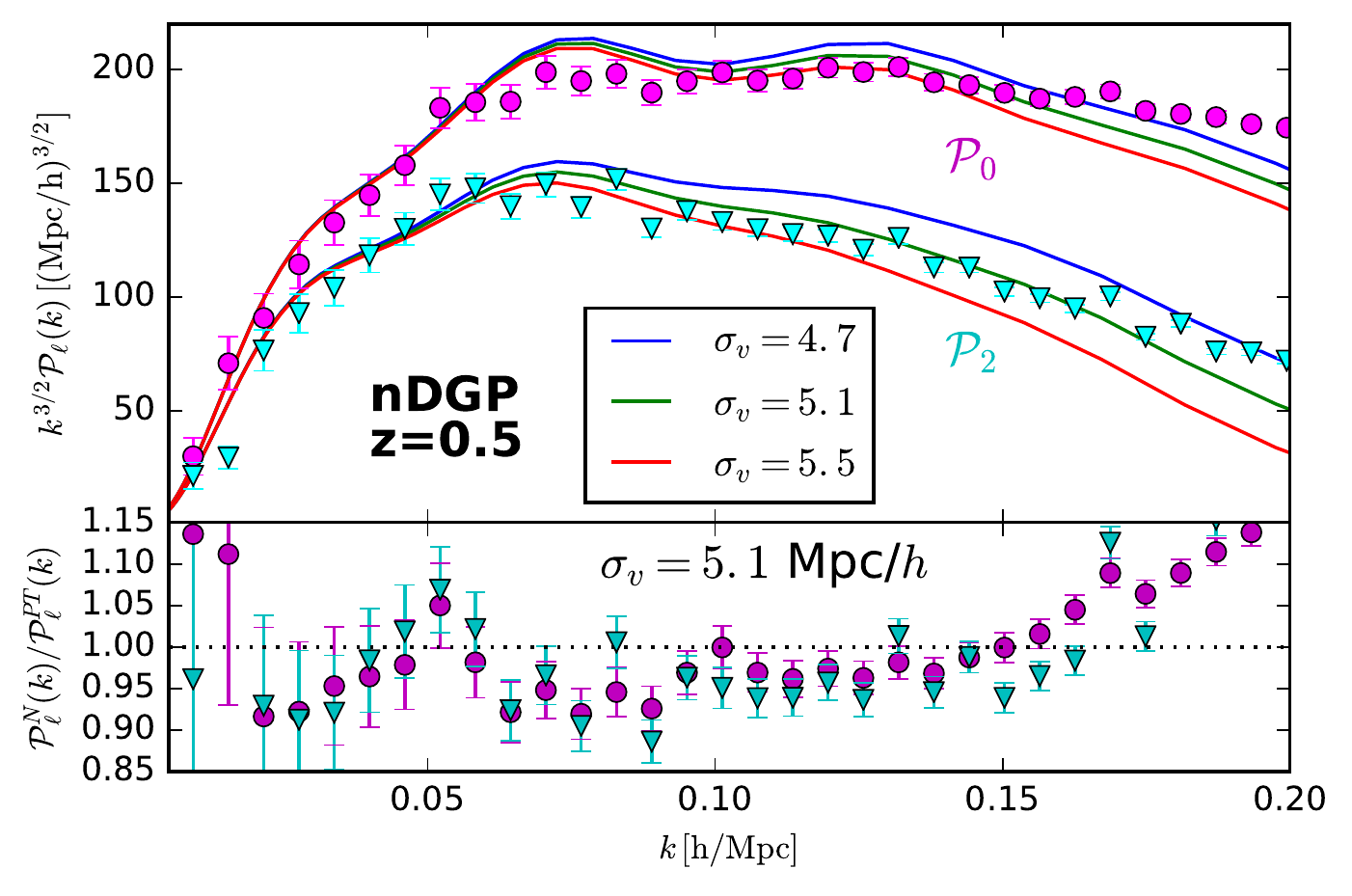}} 
  \caption[CONVERGENCE ]{ LEFT: Comparison of N-body measurements of the auto matter (blue), matter-velocity divergence (green) and auto velocity divergence (red) power spectra in real space at $z=0.5$ for nDGP. The top panels show the RegPT (solid) and SPT (dashed) power spectra multiplied by $k^{3/2}$ and the bottom panels show the fractional difference between N-body and RegPT predictions with range of $1\%$ deviation indicated by a solid arrow.  RIGHT: Comparison of N-body measurements of the redshift space monopole (magenta) and quadrupole (cyan) power spectra at $z=0.5$ for nDGP. The top panels show the multipoles multiplied by $k^{3/2}$ calculated for three values of $\sigma_v$ and the bottom panels show the fractional difference between N-body and TNS predictions for the best fit $\sigma_v$. The nDGP parameter is $\Omega_{rc}=0.438$.}
\label{pab1}
\end{figure}

To get a good benchmark for the accuracy of the GSM predictions we consider the FT of eq.(\ref{redshiftps}). By fitting $\sigma_v$ to N-body simulations we are able to accurately reproduce quasi non-linear effects which are then transferred to the correlation function. Using the RegPT prescription we are not punished by divergences in the integration over higher $\bfk$ modes. We begin by finding the best fit $\sigma_v$ and to do this we first must determine the range of validity of SPT. The left pane of Fig.\ref{pab1} shows the real space matter-matter (blue), matter-velocity divergence (green) and velocity divergence(red) power spectra modelled using SPT (dashed) and RegPT (solid) against N-body data for the nDGP model of gravity.  The $k_{\rm max}h$/Mpc  we use for the fitting of $\sigma_v$ in the multipoles is given by the solid arrow which delimits the $1\%$ deviation region. We have fitted Gaussian error bars to the data assuming a survey volume of $1 \mbox{Gpc}^3/h^3$.  
\newline
\newline
With a range of validity we can now fit the TNS free parameter $\sigma_v$. We consider the multipoles of eq.(\ref{redshiftps}) given by eq.(\ref{multipolesF}). The monopole and quadrupole, $\ell = 0,2$ respectively, are then fit up to the $k_{\rm max}$ found previously. Higher order multipoles have a very low signal to noise ratio making them problematic to measure in practice and so we will not consider them in our results. 
\newline
\newline
The right pane of Fig.\ref{pab1} shows the monopole (magenta) and quadrupole (cyan) N-body measurements against the RegPT-TNS predictions for three different values of $\sigma_v$. The fractional difference of the best fit $\sigma_v$ with N-body is shown in the bottom panels. The best fit values for $\sigma_v$ is found to be $5.1$Mpc/$h$. The best fit value for $f(R)$ and GR were found to be $6$Mpc/$h$ and $4.75$Mpc/$h$ respecitvely in \cite{Taruya:2014faa}.
\newpage
\renewcommand{\bibname}{References}
\bibliography{mybib}{}
\end{document}